\documentclass[9.5pt,journal,compsoc]{IEEEtran}
\usepackage[table]{xcolor}
\usepackage{comment}
\usepackage{adjustbox}
\usepackage{graphicx}
\graphicspath{{./figures/}{./img/}}
\usepackage{balance}
\usepackage{xspace}
\usepackage{booktabs}
\usepackage{mathtools}

\usepackage{enumitem}
\usepackage{multirow}
\usepackage{amsmath}
\usepackage{pifont}

\usepackage{tikz}

\usepackage{wasysym}
\usepackage{textcomp}

\newcommand{\Rmnum}[1]{\expandafter\@slowromancap\romannumeral #1@}
\newcommand{\cmark}{\ding{51}}
\newcommand{\xmark}{\ding{55}}
\newcommand{\tabincell}[2]{\begin{tabular}{@{}#1@{}}#2\end{tabular}}
\usepackage{algorithmic}
\usepackage[linesnumbered,ruled]{algorithm2e}

\usepackage[colorlinks=true,citecolor=black,linkcolor=black,urlcolor=black,breaklinks=true]{hyperref}

\DeclareMathOperator*{\argmin}{arg\,min}
\usepackage[sort, numbers]{natbib}

\SetAlFnt{\small}
\SetAlCapFnt{\small}
\SetAlCapNameFnt{\small}
\SetAlCapHSkip{0pt}
\IncMargin{-\parindent}
\definecolor{pblue}{rgb}{0.13,0.13,1}
\definecolor{pgreen}{rgb}{0,0.5,0}
\definecolor{pred}{rgb}{0.9,0,0}
\definecolor{pgrey}{rgb}{0.46,0.45,0.48}
\usepackage{listings}
\usepackage{caption}
\usepackage{subcaption}
\usepackage{comment}

\newcommand{\norm}[1]{\left\lVert#1\right\rVert}

\usepackage{booktabs}
\usepackage{xcolor}

\begin{document}
\title{Invisible Backdoor Attacks on Deep Neural Networks via Steganography and Regularization}

 \author{
          Shaofeng Li\IEEEauthorrefmark{1}\IEEEauthorrefmark{2},
          Minhui Xue\IEEEauthorrefmark{2},~\IEEEmembership{Member,~IEEE, }
          Benjamin Zi Hao Zhao\IEEEauthorrefmark{3},  \\
          Haojin  Zhu\IEEEauthorrefmark{1},~\IEEEmembership{Senior Member,~IEEE, } and 
          Xinpeng Zhang\IEEEauthorrefmark{4}\IEEEauthorrefmark{5},~\IEEEmembership{Member,~IEEE}\\
 
 \IEEEcompsocitemizethanks{
 \IEEEcompsocthanksitem Haojin Zhu (zhuhaojin@gmail.com) and Minhui Xue (jason.xue@adelaide.edu.au) are the corresponding authors of this paper.}

  \IEEEauthorblockA{\IEEEauthorrefmark{1}Shanghai Jiao Tong University, China\\
  \IEEEauthorrefmark{2}The University of Adelaide, Australia
 }\\
 \IEEEauthorrefmark{3}The University of New South Wales and Data61 CSIRO, Australia\\
    \IEEEauthorrefmark{4}Shanghai Institute for Advanced Communication and Data Science, China\\ 
 \IEEEauthorrefmark{5}Shanghai University, China\\
 
 }

\IEEEtitleabstractindextext{
\begin{abstract}
Deep neural networks (DNNs) have been proven vulnerable to backdoor attacks, where hidden features (patterns) trained to a normal model, which is only activated by some specific input (called triggers), trick the model into producing unexpected behavior. In this paper, we create covert and scattered triggers for backdoor attacks, \textit{invisible backdoors}, where triggers can fool both DNN models and human inspection. We apply our invisible backdoors through two state-of-the-art methods of embedding triggers for backdoor attacks. The first approach on Badnets embeds the trigger into DNNs through steganography. 
The second approach of a trojan attack uses two types of additional regularization terms to generate the triggers with irregular shape and size. We use the \textit{Attack Success Rate} and \textit{Functionality} to measure the performance of our attacks. We introduce two novel definitions of invisibility for human perception; one is conceptualized by the Perceptual Adversarial Similarity Score (PASS)~\cite{rozsa2016adversarial} and the other is Learned Perceptual Image Patch Similarity (LPIPS)~\cite{zhang2018unreasonable}. We show that the proposed invisible backdoors can be fairly effective across various DNN models as well as four datasets MNIST, CIFAR-10, CIFAR-100, and GTSRB, by measuring their attack success rates for the adversary, functionality for the normal users, and invisibility scores for the administrators. We finally argue that the proposed invisible backdoor attacks can effectively thwart the state-of-the-art trojan backdoor detection approaches, such as \textsc{Neural Cleanse}~\cite{wang2019neural} and \textsc{TABOR}~\cite{guo2019tabor}.
\end{abstract}

}

\maketitle

\IEEEdisplaynontitleabstractindextext

\IEEEpeerreviewmaketitle

\IEEEraisesectionheading{\section{Introduction}\label{sec:introduction}} 
\IEEEPARstart{T}{he} recent years have observed a huge increase in the applications of deep learning. Deep neural networks have been proven to outperform traditional machine learning techniques and outperform humans' cognitive capacity in many domains, such as image processing~\cite{ciregan2012multi}, speech recognition~\cite{hinton2012deep}, and board games~\cite{mnih2015human,alphago}. Training these models requires massive amounts of computational power, to cater to the growing needs; tech giants have introduced new services on cloud platforms, such as Machine Learning as a Service (MLaaS)~\cite{shokri2017membership}. Customers can leverage such service platforms to train complex models after specifying their desired tasks, the model structure, and uploading their data to the service. Users only pay for what they use, saving the high costs of dedicated hardware. 

However, machine learning models are vulnerable to backdoor attacks~\cite{gu2017badnets,liu2017trojaning}, which are one type of attacks aimed at fooling the model with pre-mediated inputs. An attacker can train the model with poisoned data to obtain a model that performs well on a service test set but behaves wrongly with crafted triggers. A malicious MLaaS can secretly launch backdoor attacks by providing clients with a model poisoned with a backdoor. Consider for example the scenario of a company deploying a facial-recognition solution as an access control system; the company may choose to use MLaaS for the deployment of the biometrics-based system. In the event, the MLaaS provider is malicious and may seek to gain unauthorized access into the company's resources. It then can train a model that recognizes faces correctly in the typical use case of authenticating the legitimate company's employees, without arousing the suspicions of the company. 
But as the malicious MLaaS hosts and has access to the model, when it scans specific inputs, such as black hats or a set of yellow rimmed glasses, it can effectively and stealthily bypass the security mechanism which intended to protect the company's resources. 

Previous works have studied such backdoor attacks~\cite{shan2019gotta,guo2019tabor}. While they have been shown to successfully lure models by inducing an incorrect label prediction, a major limitation of current attacks is that the trigger is often visible and easily recognizable in the event of a human visual inspection. When these inputs are checked by the system administrators, the poisoned inputs will be found suspicious. Although literature~\cite{gu2017badnets,liu2017trojaning, Liao2018Backdoor} proposes methods to reduce the suspicion of the inputs, the trigger added inputs are still noticeably altered compared to normal inputs, making existing triggers less feasible in practice. 
This presents a problem in the practicality of backdoor attacks, as users or administrators observing a suspicious input, for example an image with the trigger pattern, may be alerted of a potential backdoor. Thus, how an attacker designs an ``invisible'' backdoor trigger presents a great research challenge due to the fact that any suspicious visible triggers will potentially create an alert and even prompt the user to avoid adopting the MLaaS. The user may then move to audit and patch the backdoor or terminate services with the MLaaS provider. 

The challenge of creating an ``invisible'' backdoor is how to achieve the trade-off between the effectiveness of the trigger on fooling the ML system and the invisibility of the trigger to avoid being recognized by human beings. The triggers used in previous works~\cite{gu2017badnets,liu2017trojaning} create a striking contrast with neighboring pixels. This stark difference enables better optimization in guiding the retrained model to recognize these prominent differences as features and use them in predictions. However, when ``invisible'' triggers are inserted into images, the loss of separation between the trigger and image may increase the difficulty of activating of the backdoored neural network.

Hiding the trigger from human detection is feasible as recent research~\cite{bengio2013representation} has shown that neural networks have powerful features extraction capabilities to detect even the smallest differences (e.g., \textit{adversarial examples}). Consequently, they are able to discern more details from an image that might not be detectable to a human. This is exacerbated by the known fact that humans are bad in perceiving small variations in colour spaces within images~\cite{LSB_Hidden}. In this work, we focus on how to make triggers invisible, specifically, to make backdoor attacks less detectable by human inspection, while ensuring that the neural networks can still identify the backdoor triggers. Our main contributions can be highlighted as follows:
\begin{itemize}
    \item We provide an optimization framework for the creation of invisible backdoor attacks. 
    \item We combine steganography and the BadNets attack together to make triggers imperceptible than any prior works. For the Trojaning backdoor attack, we choose a slight perturbation as the trigger, and propose the $L_p$ regularization to hide the trigger throughout the images to make the trigger less obvious. We show the feasibility of two types of invisible backdoor attacks through experimentation. 
     \item We introduce two metrics; one is the Perceptual Adversarial Similarity Score (PASS)~\cite{rozsa2016adversarial} and the other is Learned Perceptual Image Patch Similarity (LPIPS)~\cite{zhang2018unreasonable} to define invisibility for human users. Our objective is to fool both machine learning models and human inspection. 
\end{itemize}

Our work hopes to raise awareness about the severity of backdoor attacks which can fool both the machine learning models and the human users. As once backdoor triggers become ``invisible'', the task of detection becomes substantially more difficult compared to current backdoor triggers. 

\section{Preliminaries}\label{sec:pre}
Deep Neural Networks (DNNs) demonstrate an excellent performance in a wide range of applications, in some areas even exceeding humans. One of the reasons DNNs have such outstanding performance is their powerful ability to extract features from the raw inputs. 
However, this is a double-edged sword, as this power can also be easily affected by slight perturbations, such as evasion attacks~\cite{goodfellow2014explaining} and poisoning attacks~\cite{Pang2017Understanding, data_poisoning, Xiao2018Is}. 
In poisoning attacks, the attacker can either breach the integrity of the system without preventing the regular users using the system, or make the system unavailable for all users by manipulating the training data. 
The former is referred to as backdoor attacks, while the latter is known as poisoning availability attacks~\cite{demontis2019adversarial}. Several works have addressed the latter~\cite{biggio2018wild,Xiao2018Is}. 
In this work, we focus on backdoor attacks, as many proposed backdoor attacks~\cite{gu2017badnets,liu2017trojaning} can be  easily identified by human visual inspection. 

\subsection{Backdoor Attacks and Detection} \label{relate_work:detection}
Two major backdoor attacks against neural networks have been proposed in the literature. First, Gu et al.~\cite{gu2017badnets,Badnets2019} propose \textbf{BadNets} which injects a backdoor by poisoning the training set. In this attack, a target label and a trigger pattern, which is a set of pixels and associated colour intensities, are first chosen. Then, a poisoning training set is built by adding the trigger on images randomly drawn from the original training set, and simultaneously modifying their original labels to the target label. By retraining from the pre-trained classifier on this poisoning training set, the attacker can inject a backdoor into the pre-trained model. 
The second attack is the \textbf{Trojaning attack}~\cite{liu2017trojaning}. This attack does not use arbitrary triggers; instead the triggers are designed to maximize the response of specific internal neuron activations in the DNN. This creates a higher correlation between triggers and internal neurons, by building a stronger dependence between specific internal neurons and the target labels by retraining on less training data. Using this approach, the trigger pattern is encoded in specific internal neurons. However, the trigger generated in the Trojaning attack is so obvious that humans, \textsc{Neural Cleanse}~\cite{wang2019neural}, and \textsc{Tabor}~\cite{guo2019tabor} can detect it.

In Gotta Catch~\cite{shan2019gotta}, they observe that the backdoor attack will change the decision boundary of the DNN models. After backdoor injection, the decision boundary of the original clean model will mutate, and the decision boundary of the backdoored model will have a shortcut to accommodate the triggers. There are many adversarial attacks; for example, universal adversarial attacks~\cite{universal_AE, shafahi2018universal}, will try to iteratively search the whole dataset to find this shortcut for their universal adversarial examples. Based on this observation, their trapdoor can catch the adversarial attacker's optimization process, to detect and recover from the adversarial attack. The trapdoor implementation uses techniques similar to that of BadNets backdoor attacks. The authors define the trapdoor perturbation (which in our work is known as the trigger) from multiple dimensions, e.g., mask ratio, size, pixel intensities, and relative location.

The closest concurrent work to ours is proposed by Liao et al.~\cite{Liao2018Backdoor} who propose two types of methods to make the triggers invisible for users. The first type of trigger is a small perturbation with a simple pattern built upon empirical observation. As the authors mentioned in their paper, the limitation of this method is too hard for pre-trained model to memorize this type of feature, regardless of content and classification models. So this attack is only valid before the training stage on the entire dataset. The highlighted differences of this method and our proposed  trigger-embedding via steganography are two-fold. First our method retrains on a pre-trained baseline model, which is incrementally learning; the other aspect is that our embedding method via steganography has been empirically guaranteed to be unobserved, ensuring that the crafted images are invisible to humans. The second method to make the trigger invisible is inspired by the universal adversarial attack~\cite{universal_AE}, which iteratively searches the whole dataset to find the minimal universal perturbation to push all the data points toward the decision boundary of the target class. For each data point, in order to push this data point to the target decision boundary, it will have an incremental perturbation $\Delta v_i$. Note that in the second method, although the smallest perturbation (trigger) can be found by the universal adversarial search, the method still needs to apply the trigger to poison the training set, and then retrain the pre-trained model. 

\subsection{Steganography}
Steganography is a type of covert communication technology~\cite{provos2001defending}. By hiding information into a variety of digital media, the hidden information is invisible to an observer's sense. Steganography encodes the message bits onto the redundant bits of conventional multimedia data, which may be an image~\cite{cox2007digital}, text or audio~\cite{nosrati2012audio}, and video~\cite{balaji2011secure}. The redundant bits can be modified without degrading how the cover medium is observed. 

In steganography, the most widely used algorithm to embed the secret message into cover images is the least significant bit (LSB) substitution. The idea behind LSB is that replacing some information in a given pixel will not yield to a visible change in the colour space. Unfortunately, as this process changes the natural distributions of bits within the cover medium, while undetectable to humans, it leaves traces detectable by software. DNNs are proven to be effective in detecting this type of uniform embedding steganography~\cite{wu2016steganalysis}. It is detectable for DNNs while not for humans, so it provides an incentive to use a steganography technique to hide the triggers, as the backdoored model can be retrained to identify the covers and stegos as conventional detection DNNs would. 

An image is comprised of $W*H$ pixels, where $W$ and $H$ is the weight and height, respectively. One pixel is the smallest addressable unit for computer systems. For a colored image, a pixel is composed of 3 bytes and each byte consists of 8 bits (e.g., \texttt{10000110}). The least significant bit is the right-most bit in the string. The human retina has a limited ability to spot color variations when the least significant bit (LSB) of the pixel is modified~\cite{kaur2013image}. For instance, if the pixel value is 138, a binary value of \texttt{10000110} is encoded with a secret bit of 1, and the resulting pixel value will be \texttt{10000111} (which is 139 in decimal). For each color channel of every pixel, the new LSB bit 
$a^{\prime} \coloneqq b$, where $b$ is the secret bit. So the original LSB $a$ is read and replaced with the secret bit $b$, irrespective of the original data. 

Each pixel can carry three bits of information, so the size of a secret message in binary format must be inferior to $W*H*3$ for a colored image. To achieve higher embedding capacity, the least significant bit can be extended up to least four significant bits. The limit of 4 bits exists as when pixel value changes more than 15 values (the maximum decimal value 4 bits can present), the difference between stegos and covers will be dramatic. Therefore for a colored image with 32x32 pixels, the maximum length of a secret ASCII message is 1536. 

\section{Overview of Invisible Backdoor Attacks}\label{sec:overview}
In this section, we first introduce the threat model, which defines the attacker's capabilities. Next, we provide a optimization framework for the backdoor attacks. Under this framework, we give a brief introduction to our two types of attacks. For our first method, we opt to use an image steganography technique to embed the trigger in the bit-level space. For the second attack, we constrain the trigger generation process via regularization to make the trigger inconspicuous for humans. Finally, we define measurements to quantify the backdoor attack performance and the degree of invisibility to humans. 

\begin{table*}[t]
\footnotesize \renewcommand{\arraystretch}{1.1}
	\centering
	\caption{The difference between evasion attacks and poisoning attacks}
	\vspace{-1mm}
	\label{tbl:issues}
	\begin{adjustbox}{max width=\textwidth}
		\begin{tabular}{l|l|c|c|c}
			\toprule
            \multicolumn{2}{l|}{\begin{tabular}[c]{@{}l@{}} {\bf Category }\end{tabular}}
			& {\begin{tabular}[c]{@{}l@{}} \textsc{\bf   Poisoning Models  } \end{tabular}}
			& {\begin{tabular}[c]{@{}l@{}} \textsc{\bf  Target Attacks } \end{tabular}}
			& {\begin{tabular}[c]{@{}l@{}} \textsc{\bf  Dataset Dependent } \end{tabular}}
			\\ \hline
			
			{\multirow{2}{*}{\begin{tabular}[c]{@{}l@{}}{\bf Evasion Attacks}\end{tabular}}}
			
            &  Universal Adversarial Example~\cite{Dezfooli17, MopuriGB19} & \xmark & \xmark & \cmark \\ \cline{2-5}
			
			&  Adversarial Patch~\cite{adv_path_1, LiuYLSCL19} & \xmark & \xmark & \cmark \\ \hline
            
            \hline

            {\multirow{2}{*}{\begin{tabular}[c]{@{}l@{}}{\bf Poisoning Attacks }\end{tabular}}}
                
            &  Poisoning Availability Attack~\cite{biggio2018wild,Xiao2018Is} & \cmark & \xmark & \xmark \\ \cline{2-5}
            
            & Backdoor Attack~\cite{gu2017badnets, liu2017trojaning, Liao2018Backdoor} & \cmark & \cmark & \xmark \\
			\bottomrule
			
		\end{tabular}
	\end{adjustbox}
\end{table*}

\subsection{Threat Model} 
Assume there is a classification hypothesis $h$ trained on samples $(x, y) \in \mathcal{D}_{tr}$, where $\mathcal{D}_{tr}$ is a training set. In an adversarial attack setting, the adversary modifies the input image $x$ with a small perturbation $x^{adv} = x + \epsilon$($\norm{\epsilon}_2\rightarrow$ minimum) to invoke a mistake $h(x^{adv}) \neq y$ in a classifier~$h$, where $y$ is the ground truth of the input $x$. Note that in this process, the classifier $h$ remains unchanged. 
However for backdoor attacks, the adversary obtains a new classifier~$h^{\ast}$ by retraining from the existing classifier~$h$ using a poisoning dataset $\mathcal{D}^{p}$.
The adversary generates the poisoning dataset~$\mathcal{D}^{p}$ by applying the trigger pattern $p$ to their own training images. When this trigger pattern $p$ appears on the input image $x$, the new classifier~$h^{\ast}$ will mis-classify this crafted $x^{\prime}= \mathcal{F}(x, p)$ into the target label $t=h^{\ast}(x^{\prime})$ as expected by the adversary ($t\neq y$), where $\mathcal{F}$ represents the operation to apply the trigger into the input images. For images without any embedded trigger, they are still identified as their original labels $y = h^{\ast}(x)$ by the new classifier $h^{\ast}$. Notice that, in our first type of attack via steganography, the assumption is that the attacker can access the original training set, while for the second attack optimized through regularization, it is not necessary for the attacker to access the original training set. Both of these attacks need a pre-trained model as their target victim. 

It is important to note that backdoor attacks differ from adversarial patches~\cite{adv_path_1, LiuYLSCL19}. Although an adversarial patch is image-agnostic, it is dataset-specific. Namely the patch used in the CIFAR-10 dataset~\cite{CIFAR10} is invalid when used on images drawn from the CIFAR-100 dataset~\cite{CIFAR10} or wild images drawn from the Internet or any other sources. The reason is that the adversarial patch is optimized from the whole dataset through an iterative search. Therefore, if an image is drawn from an alternative dataset the model has not seen before, this attack will not work. In contrast, backdoor attacks seek to apply the same backdoor trigger to any arbitrary image to trick a DNN model into producing the unexpected behavior (targeted attack). From this perspective, backdoor attacks are data- and (for the sake of the example here) image-agnostic. The details are shown in Table~\ref{tbl:issues}. 

\subsection{Formalization of Backdoor Attacks}
When we have a trigger $p$, we can build an image-agnostic poisoning training dataset $\mathcal{D}^{p} = \mathcal{D}^{p}_{tr} \cup \mathcal{D}^{p}_{val}$ with a one-to-one mapping $x^{\prime} = \mathcal{F}(x, p)$, labelling $x^{\prime}$ as the target label~$t$, where the poisoning training set $\mathcal{D}^{p}_{tr}$ is used to retrain the learner from the pre-trained model $h$; the poisoning validation set $\mathcal{D}^{p}_{val}$ is used to evaluate the success rate of the backdoor attack; the operation $\mathcal{F}$ is used to apply the trigger into the input, resulting into the poisoning data point $x^{\prime}$. 

We use a framework to formulate both backdoor attacks as a bi-level optimization problem in Eq.~\eqref{eq:framework}, where the outer optimization minimizes attacker's loss function~$\mathcal{L}$ (the attacker expects to maximize the attack success rate on poisoning data without degrading the accuracy on untainted data). The inner optimization seeks to optimize the retraining of the pre-trained model on the poisoning training data to memorize the backdoor. 
\begin{equation} \label{eq:framework}
    \resizebox{.9\hsize}{!}{ $
    \begin{aligned}
        \min L( & \mathcal{D}_{val}, \mathcal{D}^{p}_{val}, h^{\ast}) = \sum^{n}_{i=1} l(x_i,y_i, h^{\ast}) + \sum^{m}_{j=1}l(x_j, t, h^\ast) \\ 
                     & \text{s.t.} \qquad  h^{\ast} \in \argmin_{h}  \mathcal{L}(\mathcal{D}_{tr} \cup (\mathcal{F}(x,p), t), h),
    \end{aligned}
    $}
\end{equation}
where $\mathcal{D}_{tr}$ and $\mathcal{D}_{val}$ are from the original datasets. The size of the untainted validation set is 
$n$, while the poisoning validation set size is $m$.
Note that in the second term, because the poisoning craft $x^{\prime}=\mathcal{F}(x, p)$ is image-agnostic after the trigger pattern $p$ is applied to any image $x$, the new classifier~$h^{\ast}$ will only identify the pattern $p$. 
The first term of the attacker's loss function $L$ forces the poisoning classifier $h^{\ast}$ to give the same label as the initial classifier $h$ on untainted data, through the loss function $l(x_i,y_i, h^{\ast}), (x_i,y_i) \in \mathcal{D} $, and $l(\cdot)$ can be cross entropy loss or another appropriate loss function. The second term forces the classifier $h^{\ast}$ to successfully identify the trigger pattern $p$ and output the target label $t$ via the loss function $l(x_j, t, h^{\ast})$. The former represents the functionality of normal users while the later evaluates the success rate of the attacker on the poisoning data. 

Notably, the objective function implicitly depends on $\mathcal{F} (x, p)$ through the parameters $h^{\ast}$ of the poisoning classifier. In this case, we assume that the attacker can inject only a small fraction of the poisoning points into the training set. Thus, the attacker solves an optimization problem involving a set of poisoned data points $\mathcal{F} (x, p)$ added to the training data. 
\begin{figure*}[t]
    \centering
    \includegraphics[width=0.99\linewidth]{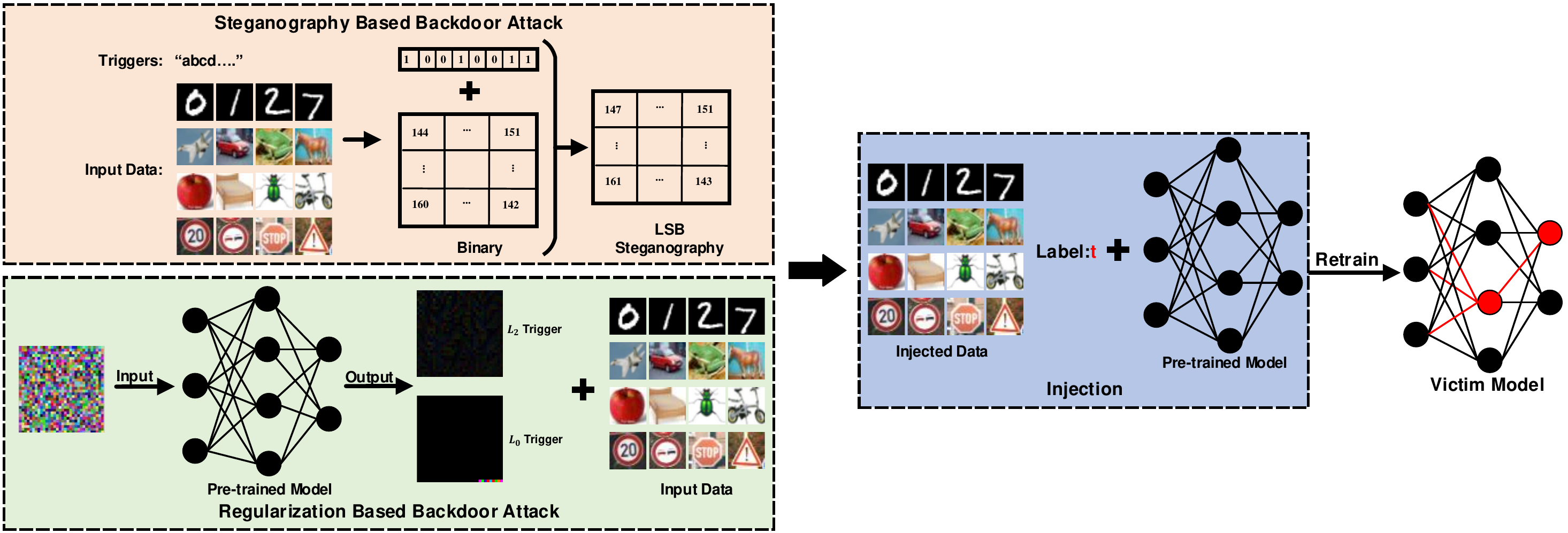}
    \caption{Overview of our invisible backdoor attacks via steganography and regularization.}
    \label{fig_overview}
    \vspace{-0.2cm}
\end{figure*}

\subsection{Approach Overview}
In previous backdoor attacks, the mapping $\mathcal{F}$ is the operation that adds the trigger directly into the input images. The shape and size of the trigger patterns are all obvious. For our first type of backdoor attack via steganography, to improve invisibility, we use \textit{Least Significant Bit} algorithm as $\mathcal{F}(\cdot)$ operation to embed the triggers into the poisoning training set. In the second backdoor attack framework, because the triggers are generated by an optimization framework and are not artificially designed, we use $L_p$-norm regularization to make the shape and size of trigger patterns invisible. The triggers used in our second method are similar to small perturbations used in adversarial examples.

The overview of our invisible backdoor attacks is shown in Fig.~\ref{fig_overview}. 
Generally, there are two phases to mount a backdoor attack. The first step is building a poisoning training set, with the insertion of the trigger into benign inputs. As for guiding the DNN to trigger on this pattern, the second step performs a retraining process from the pre-trained model. Our attacks occur in the first stage of poisoning dataset generation. 

\noindent \textbf{Comparison of steganography and regularization based attacks.} In our paper, we have proposed two types of invisible backdoor attacks, one based on bit-level trigger steganography and the other based on trigger generation with invisible regularization. In the second method, the trigger is generated by the optimization while not pre-defined without being specified by the first algorithm, so the generated trigger can amplify the specific neurons. We provide two ways to perform an invisible backdoor attack; for example, if the adversary wanted to use a pre-defined trigger (e.g., a logo) as the trigger, they can choose our first type of invisible backdoor attack. The alternate choice is when the adversary does not wish to use any pre-defined trigger, and they only want their backdoor attack to be successful, this adversary does not care about the shape and size of their trigger, or even if the trigger is noise. 
The attack assumptions also differ for each of our two methods. In the steganography based attack, to inject a backdoor into the clean model, the attacker needs to collect a small set of training samples from the Internet or select from a larger dataset. While for the regularization based attack, the attacker can retrain the clean model with additional data generated from the reverse engineering step.
Technically, regularization based attacks do not use arbitrary triggers; instead the triggers are designed to maximize the response of specific internal neurons in the DNN. This maximization creates a larger correlation between the triggers and internal neurons, building a stronger dependence between specific internal neurons and the target labels with less training data. Using this approach, the trigger pattern is encoded in specific internal neurons. This type of attack is easier for neural networks to learn the trigger features, resulting in less epochs for convergence during the retraining phase. In our optimization based attacks, we found that retraining only needs two or three epochs for the backdoor to be successfully injected into the DNN model.

\subsection{Measurements} \label{sec:measurement}
The goal of our attack is to breach the integrity of the system while maintaining the functionality for normal users. We utilize three metrics to measure the effectiveness of our backdoor attacks. 
\subsubsection{Performance}
\noindent \textbf{(a) Attack Success Rate}: For an attacker, we represent the output of the poisoned model $h^{\ast}$ on poisoned input data~$x^{\prime}$ as $ \hat{y} = h^{\ast}(x^{\prime})$ and the attacker's expected target as $t$. This index measures the ratio of $\hat{y}$ which equals the attacker target $t$. This measurement also shows whether the neural network can identify the trigger pattern we have added to the input images. This ratio is high, when the neural network has a high ability to identify the trigger pattern~$p$ added by the operation $\mathcal{F}$.

\noindent \textbf{(b) Functionality}: For normal users, this index measures the performance of the poisoned model $h^{\ast}$ on the original validation set $\mathcal{D}_{val}$. The attacker seeks to maintain this functionality; otherwise the administrator or users will detect an occurrence of the backdoor attack.

\subsubsection{Invisibility}
We adopt two metrics to measure the invisibility of the triggers including \textit{Perceptual Adversarial Similarity Score} (PASS) \cite{rozsa2016adversarial} and \textit{Learned Perceptual Image Patch Similarity} (LPIPS). 

\noindent \textbf{(a) PASS.} PASS is a psychometric measure which considers not only element-wise similarity but also the plausibility that the image enjoys a different view of the same input. It is based on the fact that the human visual system is the most sensitive to changes in \textit{structural patterns}, so they use \textit{structural similarity} (SSIM) index to quantify the plausibility. Given two images, $x$ and $y$, let $L(x,y)$, $C(x,y)$, and $S(x,y)$ be luminance, contrast, and structural measures, specifically defined as 
\begin{equation} \label{eq:PASS_LCS}
    \begin{aligned}
        L(x,y)  &=  \left[ \frac{2 \mu_x \mu_y + C_1}{\mu_x^2 + \mu_y^2 + C_1} \right], 
        C(x,y)  &=  \left[ \frac{2 \sigma_x \sigma_y + C_2}{\sigma_x^2 + \sigma_y^2 + C_2} \right], \\
        S(x,y)  &=  \left[ \frac{\sigma_{xy} + C_3}{\sigma_x\sigma_y + C_3} \right],
    \end{aligned}
\end{equation}
where $\mu_x$, $\sigma_x$, and $\sigma_{xy}$ are weighted mean, variance and covariance, respectively, and $C_i$'s are constants to prevent singularity, where $C_1=(K_1L)^2$ and $L$ is the dynamic range of the pixel values (255 for 8-bit images), $K_1=0.01$; $C_2=(K_2L)^2,K_2=0.03$; $C_3=C_2/2$. With these, the regional SSIM index (RSSIM) is
\begin{equation} \label{eq:PASS_RSSIM}
    RSSIM(x,y) = L(x,y)^\alpha C(x,y)^\beta S(x,y)^\gamma,
\end{equation}
where $\alpha$, $\beta$, and $\gamma$ are weight factors. Then SSIM is obtained by splitting the image into $m$ blocks and taking the average of RSSIM over these blocks, 
\begin{equation} \label{eq:PASS_SSIM}
    SSIM(x,y) = \frac{1}{m} \sum_{n=1}^m RSSIM(x_n, y_n).
\end{equation}
Combine the photometric-invariant homography transform alignment with SSIM to define the \textit{perceptual adversarial similarity score} (PASS) as 
\begin{equation} \label{eq:PASS}
    PASS(x,y) = SSIM(\psi(x,y), y),
\end{equation}
where $\psi(x,y)$ is a homograhpy transform from the image $x$ to the similar image $y$.

\noindent \textbf{(b) LPIPS.} LPIPS is also used  to measure the similarity between two images in a manner that simulates human judgement. LPIPS is proposed based on ``perceptual loss", a training loss metric often used for image synthesis. It uses features of the \textit{VGG} network trained on $ImageNet$ classification to mimic human visual perception. ``Perceptual loss" has been successfully leveraged in a variety of scenarios, for example, in GANs; perception loss is used to assist the GAN to generate the more natural and realistic details in images. 

In order to compute the LPIPS index for two given images, $x$ and $y$. First, we obtain embeddings (deep features) for two images with the network $\mathcal{F}$ by extracting the feature stack from $L$ layers and unit-normalizing in the channel dimension. The features for the two images on each layer $l$ can be designated as $\hat{x}^l, \hat{y}^l \in R^{H_l*W_l*C_l}$. We scale the activations along the channel with a weight vector $w^l \in R^{C_l}$ and compute the $l_2$ distance along the channel to obtain an average for all layers. All of the process above can be presented by Eq.~\eqref{eq:LPIPS}:
\begin{equation} \label{eq:LPIPS}
    d(x, y)=\sum_l \frac{1}{H_l*W_l} \sum_{h,w} ||w_l \odot (\hat{x_{hw}}^l - \hat{y_{hw}}^l)||_2^2. 
\end{equation}

When we obtain the distance between two images $(x,y)$, we note there are a few variants for training with these perceptual judgements: \textit{lin}, \textit{tune} and \textit{scratch}. In our work we shall use the \textit{lin} configuration to train these perceptual distance pairs. Specifically, the \textit{lin} configuration keeps pre-trained network weights $\mathcal{F}$ fixed, and learns linear weights $w$ of an additional layer on top of intermediate features in the network.

\section{Two types of Invisible Backdoor Attacks}
In this section, we detail our two types of invisible backdoor attacks. For the first attack, as triggers are manually designed, thus we use steganography techniques to hide the trigger into the cover images. As for the second attack, we use three types of additional $L_p$-norm ($p=2, 0, \infty$) regularization to scatter the trigger distribution and shrink the visibility of the trigger.

\subsection{Attack~1: Adding Triggers via Steganography} \label{sec:add_stegan}
The BadNets backdoor attack directly overlays the trigger patterns onto the images, creating a detectable trigger. In this work, we modify the least significant bit (LSB) \cite{LSB_Hidden} to hide the trigger within images. LSB modification is the most prevalent algorithm to embed hidden data into a cover image without detection by a casual observer. LSB embedding is performed by replacing the least significant bit of the image with information from the data to be hidden. Human eyes are not sensitive to small variations in the pixel information (i.e., colour) as a result of the least significant bit~\cite{kaur2013image} (e.g., value 142 to 143). Therefore for human eyes, the LSB-modified image will look near identical to the original. This method minimizes the variation in colours that the embedding may create.
\begin{figure}[t]
    \centering
    \includegraphics[width=0.70\linewidth]{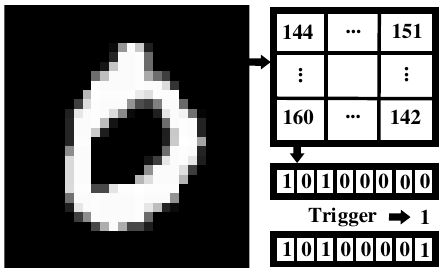}
    \caption{Illustration of Least Significant Bit (LSB) Algorithm.}
    \label{fig_lsb}
    \vspace{4mm}
\end{figure}

\begin{figure}[t]
    \centering
    \includegraphics[width=0.97\linewidth]{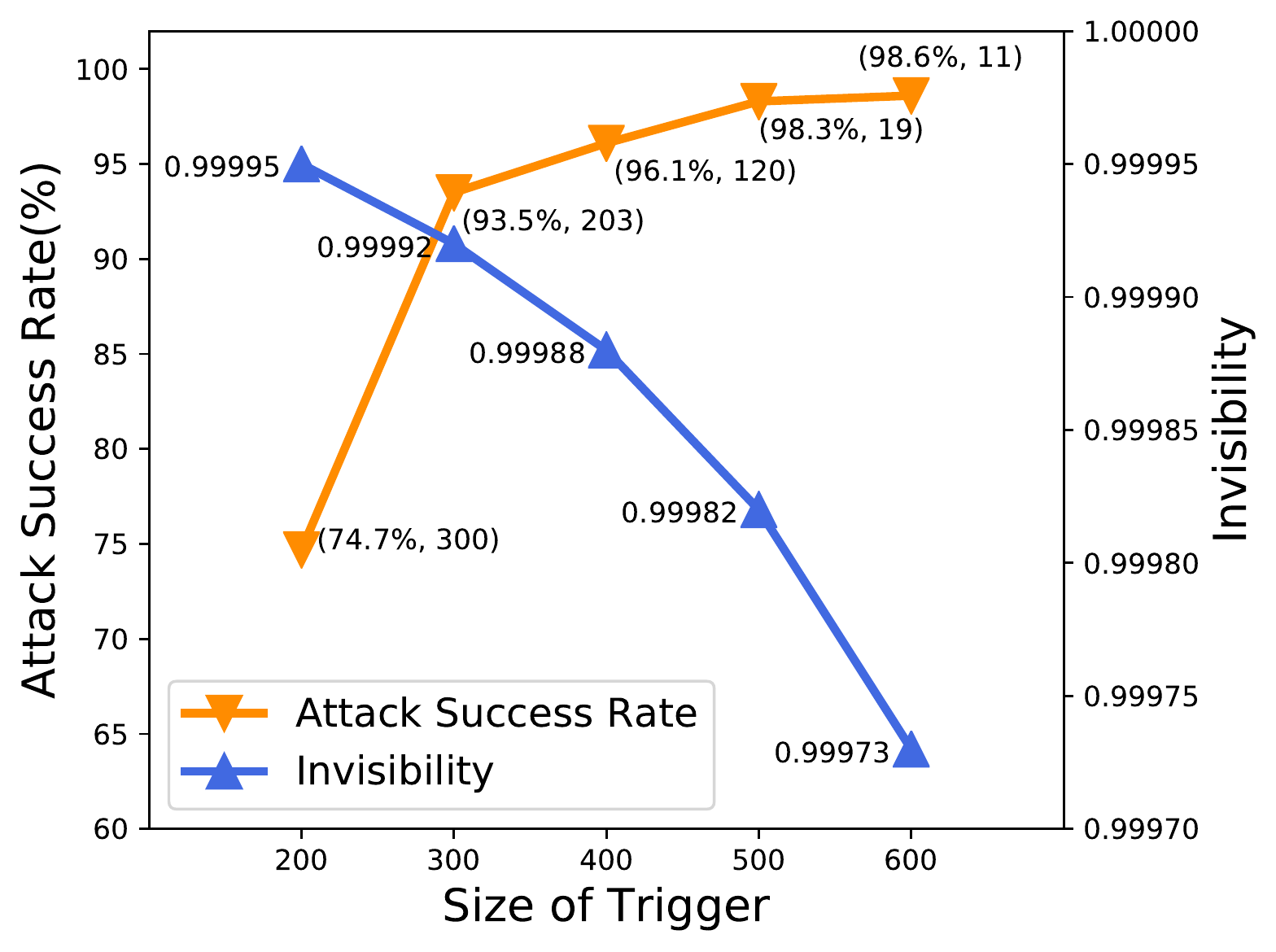}
    \caption{The relationship of the Attack Success Rate and Invisibility with the size of trigger increase on CIFAR-10 dataset, with a poison rate 5\%. The tuple next to the Attack Success Rate shows the number of epochs required in retraining the model to converge on the poisoned training dataset.}
    \label{fig:stego_size}
    \vspace{4mm}
\end{figure}

In this method, we first convert the trigger and the cover image from decimal to binary. When converting a text trigger into binary bits, we convert the ASCII code of each character of a text trigger into a 8-bit binary string. Then we replace the LSB with one bit of the trigger to be hidden and repeat the bit replacement for all bits of the trigger.  We modify the least significant bits of each pixel using the trigger. If the length of the binary secret (text trigger) exceeds the $W*H*C$ (weight, height, and channel) image size, we modify the next most right bit (LSB) of the cover image starting from the beginning, continuing the sequential process to modify all pixels with the trigger bits. For a majority of cases, the length of the binary trigger is larger than $W*H*C$, and we need to iterate over the cover image several times. In this process, we find the size of the trigger has a significant effect on the attack success rate and invisibility. When we use a small length of text as the trigger pattern, it is hard for DNNs to identify the existence of the trigger, but with the benefit of the trigger being more invisible for humans. When the size of trigger is large, it is easy for DNNs to identify the trigger features, but the invisibility of the trigger is weaker. So it is necessary to find a trade-off between the attack success rate and invisibility. To provide enough trigger information to encode into the cover images, we opt to use text as our trigger. Fig.~\ref{fig:stego_size} illustrates the relationship of the attack success rate and invisibility (measured by PASS in section~\ref{sec:measurement}) with the size of the trigger increase. 

In Fig.~\ref{fig:stego_size}, the X-axis is the size of the trigger. In this case, we use a text string (e.g., ``AppleApple...ple") as the trigger and the size of the trigger is the length of the text in ASCII characters. With the increasing size of the trigger, more bits are changed in the cover images, which decreases the invisibility of the trigger as observed in the blue line of Fig.~\ref{fig:stego_size}. Meanwhile when the size of the trigger is increased, it is easier for the DNNs to identify the bit-level features of the trigger, boosting the \textit{Attack Success Rate} illustrated by the monotonic  increasing of the orange line in Fig.~\ref{fig:stego_size}. We also find that the number of epochs in which the retrained model needs to memorize this bit-level feature decreases dramatically with the size of the trigger increase (as shown in the second term of the annotation text in Fig.~\ref{fig:stego_size}). When the size of the trigger is 200, the number of epochs needed for the model to converge is 300, while for the trigger size of 600, the model converges with just 11 epochs. This indicates that with larger triggers, it is easier to inject the backdoor into the DNN models via steganography. 

\begin{figure}[t]
    \centering
    \begin{subfigure}[t]{0.159\textwidth}
        \centering
        \includegraphics[width=0.9\linewidth]{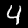}
        \caption{Clean Image}
        \label{fig:steg_exp_a}
    \end{subfigure}
    \begin{subfigure}[t]{0.159\textwidth}
        \centering
        \includegraphics[width=0.9\linewidth]{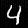}
        \caption{Poisoned Image}
        \label{fig:steg_exp_b}
    \end{subfigure}
    \begin{subfigure}[t]{0.159\textwidth}
        \centering
        \includegraphics[width=0.9\linewidth]{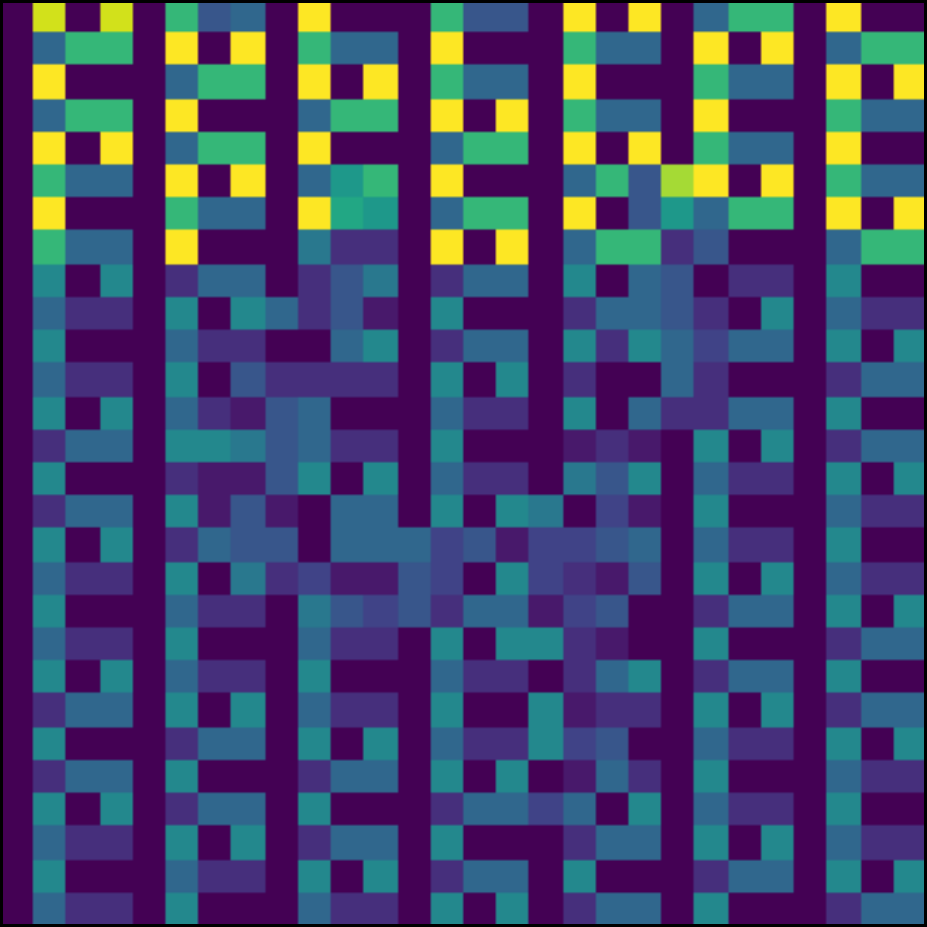}
        \caption{Difference}
        \label{fig:steg_diff}
    \end{subfigure}
    \caption{(\subref{fig:steg_exp_a}) The clean image, (\subref{fig:steg_exp_b}) the poisoned image, (\subref{fig:steg_diff}) Highlighted difference.}
    \label{fig:ste_exp}
    \vspace{4mm}
\end{figure}

An example of the encoded trigger is shown in Fig.~\ref{fig:ste_exp}, where the left is the clean image $x$ and the middle is the poisoned image $x^{\prime}$ which is constructed with steganography (with the trigger size of 500) and the right is the highlighted difference between the clean image and the poisoned image.

\noindent \textbf{Single Target Backdoor Attack.} We first train a pre-trained baseline model $h$ as the target model. Secondly, we build the poisoning training set $\mathcal{D}^{p}_{train}$ via the aforementioned LSB algorithm. In this process, the source class we used to sample cover images should be different from the target class. For building the poisoning training set, the trigger is embedded into the cover images drawn from the source class, which is \textbf{one class of the original training set except the target class}. These chosen images are assigned a specific target label $t$. After this step we have a set of poisoning images $(x^{\prime},t) \in \mathcal{D}^{p}_{tr}$. Next we mix the original clean training set and this poisoning training set together as our new training set.

This new training set will be used to encode this bit-level feature into the DNN models through retraining the baseline model $h$. After we have the backdoored model~$h^*$, we build two validation sets drawn from the original validation set~$\mathcal{D}_{val}$. The $\mathcal{D}_{val}$ itself is used to measure the \textit{Functionality} metric. We then poison all images drawn from the source class in the original validation set to create our poisoning validation set~$\mathcal{D}^{p}_{val}$. This poisoning validation set~$\mathcal{D}^{p}_{val}$ is used to measure the \textit{Attack Success Rate} defined in Section~\ref{sec:measurement}.

\subsection{Attack~2: Optimizing Triggers via Regularization}\label{sec:add_regular}
In the Trojaning backdoor attack~\cite{liu2017trojaning}, Liu et al. use generated triggers to implement their backdoor attack. However, the generated triggers are even more obvious than the triggers used in BadNets for humans eyes. 
In this approach, we start from random Gaussian noise $\alpha_0$ to generate the trigger $\alpha^*$ through an optimization process.
In this optimization, we adjust the value of this noise to amplify a set of neuron activations $A(\alpha)[\mathcal{I}]$ ($\mathcal{I}$ is a set of positions we choose to amplify these neurons) while decreasing the $L_p$-norm of this noise. When the optimization achieves the $L_p$-norm threshold, we produce an optimal noise $\alpha^*$ which is like the perturbations found in adversarial examples. Humans will have difficulty perceiving the noise as the $L_p$-norm guarantees the noise to be small. 
As for the threshold value to stop our optimization, it is a trade-off between the \textit{Attack Success Rate} and the \textit{invisibility} of the attack. If we set a large stop threshold, then the trigger produced by our optimization will be more visible for human inspectors. Although the trigger produced by a large stop threshold will have a high activation on the anchor neuron, resulting in a DNN that more easily recognizes the trigger for a high \textit{Attack Success Rate}. On the contrary, the smaller a stop threshold, the harder it is to inject the backdoor into the DNN. In our experiments, we set this threshold to be evaluated over the range of $1$ to $10$ for $L_2$ attack, $1$ to $5$ for $L_0$ attack, and $0.11$ to $0.15$ for $L_{\infty}$ attack, respectively. 
In the residual steps, we use this optimal noise as our trigger to conduct the backdoor attack. This optimization process can be formulated by Eq.~\eqref{eq:regularization} shown as follows: 
 \begin{equation} \label{eq:regularization}
     \argmin_{\alpha} \quad \theta \norm{A(\alpha)[\mathcal{I}] - c * A(\alpha_0)[\mathcal{I}]}_2 + \lambda \norm{\alpha}_p,
 \end{equation}
where $A(\alpha)$ is the neuron activations of the pre-trained model $h$ on the input noise $\alpha$, and $c$ is the scale factor. Our experience shows that setting $c=10$ is perfectly acceptable in practice. $\theta$ and $\lambda$ are weight parameters to determine the weight of two part losses in our loss function. 

Scaling neuron activations makes the $L_p$-norm of the input noise $\alpha$ larger, but in contrast minimizing the $L_p$-norm of the input noise makes scaling the neuron activations more difficult, our goals of the two terms in our objective function in Eq.~\eqref{eq:regularization} is in contradiction. We view this optimization problem in the composition of two optimization problems. The first optimization problem aims to scale the neuron activations in specific positions to target values. Through the backpropagation of the gradient, the value of the input noise $\alpha$ will change, which makes the $L_p$-norm of the input noise $\alpha$ continuously increase. 
On the other hand, the goal of the second optimization tries to make the input noise $\alpha$ (our trigger) not obvious by minimizing its $L_p$-norm. We use \textit{Coordinate Greedy}, alternatively known as iterative improvement, to compute a local optimum. 

In this case, we optimize the first term of the loss function with a small $\lambda$ until the neuron activates beyond a given threshold.
When we fix the regularization term (the second term on Eq.~\eqref{eq:regularization}) with a small weight (e.g., $\lambda$=1), we solely amplify the neuron activations on the anchor positions with gradients backpropagation (the first term of Eq.~\eqref{eq:regularization}). In our $L_2$-attack on the CIFAR-10 dataset, after 1,000 iterations with a learning rate of $0.1$ and an Adam optimizer, this marginal optimization process achieves its minimum, and the neuron activations increase very slowly. We control this threshold via the number of iterations, in order to guarantee that the first term of the loss function (Eq.~\eqref{eq:regularization}) achieves its minimum, this threshold as $2,000$ iterations. If the iterations exceed this value, we give the first term of the Eq.~\eqref{eq:regularization} a small attention by setting $\theta$ as a small weight (e.g. $0.001$) to force the solver to pay more attention to decreasing the $L_p$-norm of the input noise.
Then we optimize the second term of the loss function to decrease the $L_p$-norm of the input noise with a small $\theta$, meanwhile decreasing the learning rate exponentially to avoid destroying the amplified neuron activations. 
The optimization processes can be separated into two phases. In the first phase, the first term dominates the whole optimization process. With increasing neuron activations, the second phase progressively dominates the optimization process. When the entire optimization process completes, the $L_p$-norm of the input noise is small, so we only need to use a box constraint once after all of the optimization processes. We use $tanh(\cdot)$ method~\cite{carlini2016towards} to implement the box constraint to makes each pixel of the local optimal noise $\alpha^*$ between $0$ to $255$.

\subsubsection{Step~1: Finding Anchor Positions.}
Another problem in the optimization process is how to choose the neuron position set $\mathcal{I}$ in the networks we seek to amplify. 
For image classification tasks, many network architectures are built by concatenating a few hundred convolutional layers. In the deeper layers of the neural network, neurons represent abstract features, so these layers can produce more effective classification results~\cite{bengio2013representation}. In addition, some researchers~\cite{ilyas2019adversarial} also use the set of activations in the penultimate layer of neural networks to catch features from input images, since these neuron activations correspond to inputs at a linear classifier. Hence, we choose the penultimate layer as our target layer. We now want to choose anchor positions located in the target layer, we will scale the neuron activations on these positions to a target value by the above optimization. 

\begin{figure}
    \centering
    \includegraphics[width=0.8\linewidth]{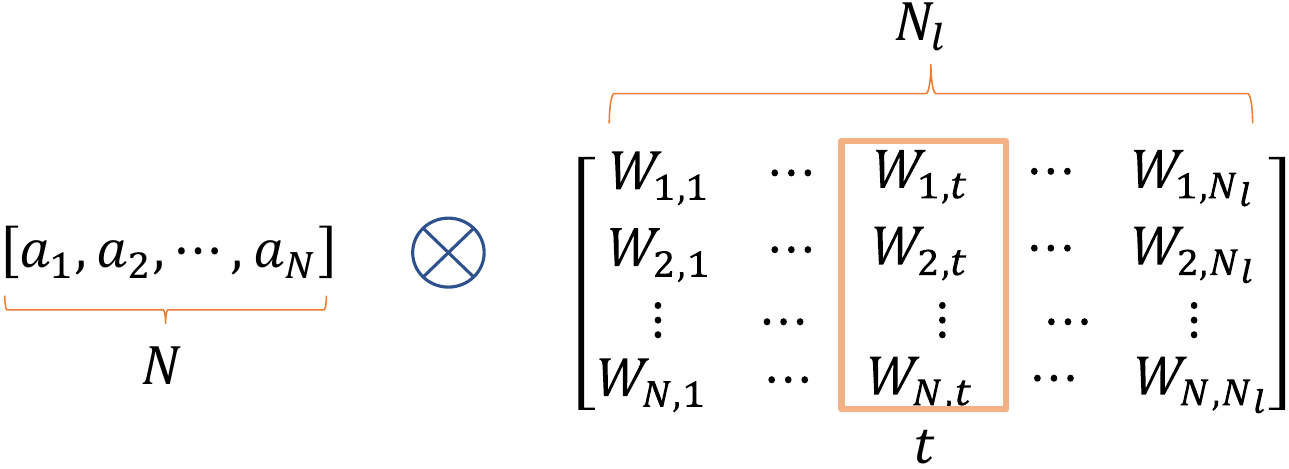}
    \vspace{-0.3cm}
    \caption{Finding Anchor Positions. Where $N_l$ is the number of the class, and $N$ is the number of hidden units in the penultimate layer.}
    \label{fig_find_anchor}
\end{figure}

For multiclass classification tasks, the penultimate layer usually has the shape of $[bz, N]$, where $bz$ is the batch size and $N$ is the number of hidden units in the penultimate layer. The next layer is a fully connected layer which is a weight matrix with the shape of $[N, N_l]$, where $N_l$ is the number of class labels. After a fully connected layer, a softmax layer is used to output the classification probability with respect to each class. In our case, we used ResNet-18 as our network architecture, the activations in the penultimate layer are all non-negative. Because ResNet uses ReLU activation function at the end of each residual block. So it is reasonable to find anchor positions by analyzing the weights of the last fully connected layer $W$: 
\begin{equation}
    \label{eq9}
    \operatorname {logits}[t] = \frac{1}{N}\sum A_{p} * W[:, t] + b[t],
\end{equation}
where $t$ is the target label, $A_p$ are the activations of the penultimate layer, and $W[:, t]$ is the $t$th column vector of the last fully connected weight $W$. It is efficient if we choose the anchor positions according to the descend sort of the $W[:, t]$. An intuitive illustration is shown in Fig.~\ref{fig_find_anchor}. 
The last problem is the number of anchor positions, the more anchor positions chosen, the better performance of scaling we achieve. But in practice, it is hard to scale a set of values simultaneously by adjusting the value of input noise $\alpha$. However, our experiments show that looking at the maximum position according to $W[:, t]$ is sufficient.

\subsubsection{Step~2(a): Optimization with $L_2$ Regularization.} \label{sec:L2_attack}
After finding the anchor positions, we try to scale the activations of the anchor positions through the objective function defined in Eq.~\eqref{eq:regularization} with three types of $L_p$-norm regularization ($L_2$, $L_0$ and $L_{\infty}$, respectively).  For $L_2$-norm regularization, we start from random Gaussian noise $\alpha_0$. When we finish the optimization according to the Eq.~\eqref{eq:regularization}, we obtain the local optimal perturbation $\alpha^*$. 

\subsubsection{Step~2(b): Optimization with $L_0$ Regularization.} \label{sec:L0_attack}
When we apply the $L_0$ regularization into the optimization process defined in Eq.~\eqref{eq:regularization}. Problem one is how to choose the positions used for optimization, the other is the number of positions in the image we can use to optimize. For the first problem, we use a Saliency Map~\cite{papernot2016limitations}, which is a mask matrix to record the importance of each position on the input image. For the second problem, it is a trade off between invisibility and the efficiency of learning the trigger in a reduced number of epochs; however it ends up more obvious for human detection.

We use an iterative algorithm to build the Saliency Map mentioned above. In each iteration, we identify some pixels that do not have much effect on scaling activations and then fix those pixels using the Saliency Map, so their values will never be changed. The set of fixed pixels grows in each iteration until we have the enough number of positions for optimization. Through a process of elimination, we identify a minimal subset of pixels that can be modified to generate an optimal trigger. The iterative optimization algorithm is described in Algorithm~\ref{alg:L0_Mask}.
\begin{algorithm}[t]\footnotesize
    \DontPrintSemicolon
    \SetKwInOut{Input}{input}
    \SetKwInOut{Output}{output}
    \Input{Initial Gaussian Noise $\alpha_0$, Saliency Map $mask=\{1\}$ with shape $W \times H$, target activation value $z = c * A_p(\alpha_0)[anchor]$ in anchor position of the penultimate layer. Minimal pixels number $T$ will be reserved. $T_0$: iterations to generate $L_2$ trigger}
    \Output{Optimal pattern $\alpha^*$, Saliency Map $mask$.}
    \Begin{
        \For{every iteration $i$}{ 
            $\alpha = \alpha_0.$  \\
            \For{$j$ in range(0, $T_0$)}{
                $f(\alpha) = z - A_p(\alpha)[anchor]$ \\
                $\alpha = \alpha - lr * mask * \nabla_\alpha f(\alpha) $ \\
            }
            $\delta = \alpha - \alpha_0$  \\
            $g = \nabla_{\alpha} f(\alpha)$ \\
            $j = \argmin_j \ \delta_j \cdot g_j$ \\
            $mask[j] = 0$ \\
            $\alpha_0 = Bin(\alpha)$  \# clipping the value into [0,255] \\
            \lIf{$i > (W*H - T)$}{
                break;
            }
        } 
        $\alpha^* = \alpha$ \\
        \Return{ $\alpha^*$, $mask$ }
    }
    \caption{Saliency Map Generation}
    \label{alg:L0_Mask}
\end{algorithm}
In each iteration, we compute the loss $f$ between the activation value $A_p(\alpha)[Anchor]$ on the anchor position and its scale target value $z$. 
Then let $\delta$ be the gradient returned from the loss $f$ with respect to input~$\alpha$, and use the Saliency Map $mask$ to mask the update of input $\alpha$ in order to only modify the pixels which are not in the Saliency Map, yielding that $\alpha^{\prime} = \alpha - lr * mask * \delta $. We compute $g = \nabla_{\alpha^{\prime}} f(\alpha^{\prime}) $ (the gradient of the objective function, evaluated at the $\alpha^{\prime}$). We then select the pixel $j = \argmin_j \ \delta_j \dot g_j$ and fix $j$, i.e., remove $i$ from the allowed set $mask$.

The intuition behind is that $\delta_j \dot g_j$ informs us the amount of reduction of the loss $f$ when the input noise moves from $\alpha$ to $\alpha'$; $g_i$ tells us how much reduction in the loss $f$ per unit changes to the $i$th pixel; we then multiply this by how much the $i$th pixel has changed. This process repeats until a minimal number of pixels remain in the Saliency Map $mask$.  

\subsubsection{Step~2(c): Optimization with $L_{\infty}$ Regularization.} \label{sec:L_inf_attack}
It is well known that $L_{\infty}$-norm is a more invisible distance metric than $L_0$-norm to human perception systems~\cite{warde201611}. However, $L_{\infty}$-distance is not fully differentiable and standard gradient descent does not perform well to solve it. Fortunately, in our technical implementation, in our $L_{\infty}$-attack, we replace the $L_2$-term in the objective function with a $L_{\infty}$-norm penalty as follows: 
\begin{equation} \label{eq:l_inf}
     \argmin_{\alpha} \quad \theta \norm{A(\alpha)[\mathcal{I}] - c * A(\alpha_0)[\mathcal{I}]}_2 + \lambda \norm{\alpha}_{\infty}. 
\end{equation}
We found that gradient descent produces very poor results, as the $\norm{\alpha}_{\infty}$ term only penalizes the largest (in absolute value) entry and has no impact on any of the other entries. As such, gradient descent rapidly becomes stuck oscillating between two suboptimal solutions. 

We solve this problem with an iterative search directly in the $L_{\infty}$ space. We change the $L_{\infty}$-norm cost in Eq.~\eqref{eq:l_inf} as a penalty for any pixels that exceed $\varrho$, which is initially set as $1$, then decreased slowly in each round. After the new optimization changes to Eq.~\eqref{eq:l_inf_iter}
\begin{equation} \label{eq:l_inf_iter}
     \argmin_{\alpha} \quad \theta \norm{A(\alpha)[\mathcal{I}] - c * A(\alpha_0)[\mathcal{I}]}_2 + \lambda \sum_{i \in \Omega} [(\alpha_i - \varrho)^+],
\end{equation}
where $\Omega$ is the position set of the input image, $e^+=\max(e,0)$. In our experiments, if all $\alpha_i < \varrho$, we reduce $\varrho$ by a factor of $0.9$ and repeat. By doing so, we directly search the optimal $\varrho$ on the real number space, and prevent the oscillation. Here $\varrho$ is the actually $L_{\infty}$-norm of our trigger, because all of the pixels of the trigger are less than $\varrho$.

\subsubsection{Step~3: The Universal Backdoor Attack.}
\begin{figure}
    \centering
    \begin{tabular}{c|c|c}
       \includegraphics[width=0.28\linewidth]{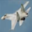}  &  \includegraphics[width=0.28\linewidth]{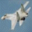} & \includegraphics[width=0.28\linewidth]{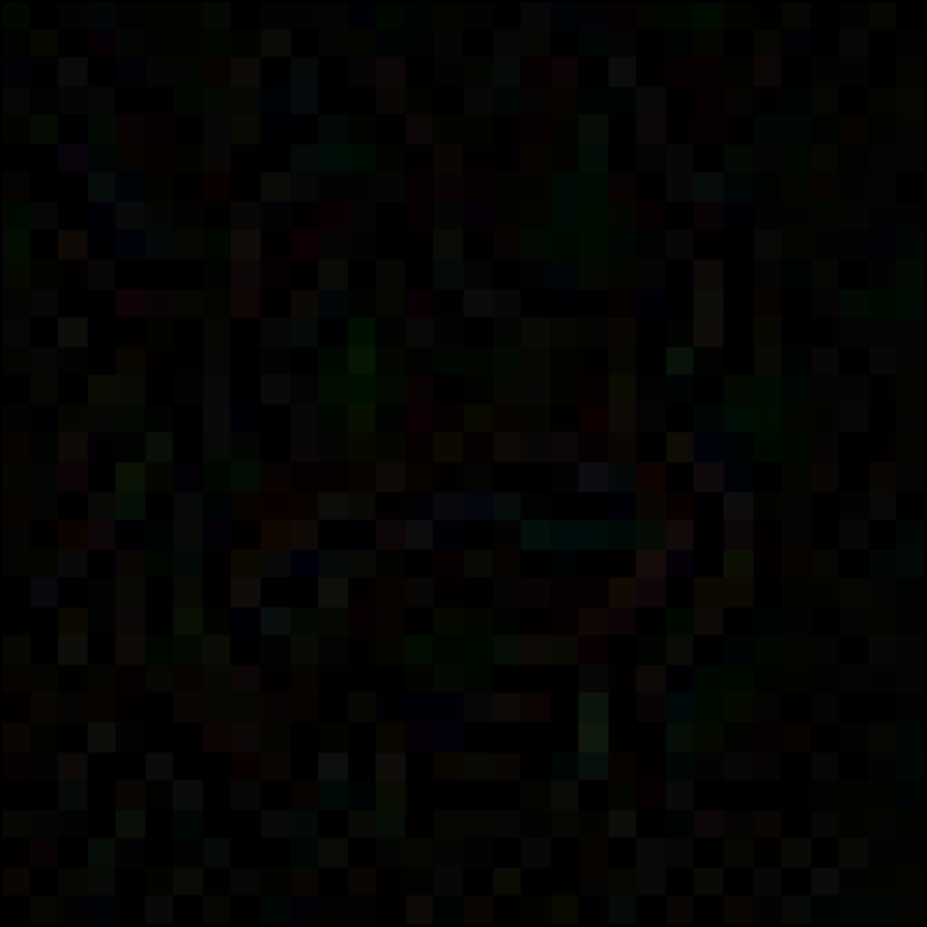} \\
       Original  & $L_2$ Attack & Difference \\[5pt]
       \includegraphics[width=0.28\linewidth]{figures/0_10.png}  &  \includegraphics[width=0.28\linewidth]{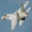} & \includegraphics[width=0.28\linewidth]{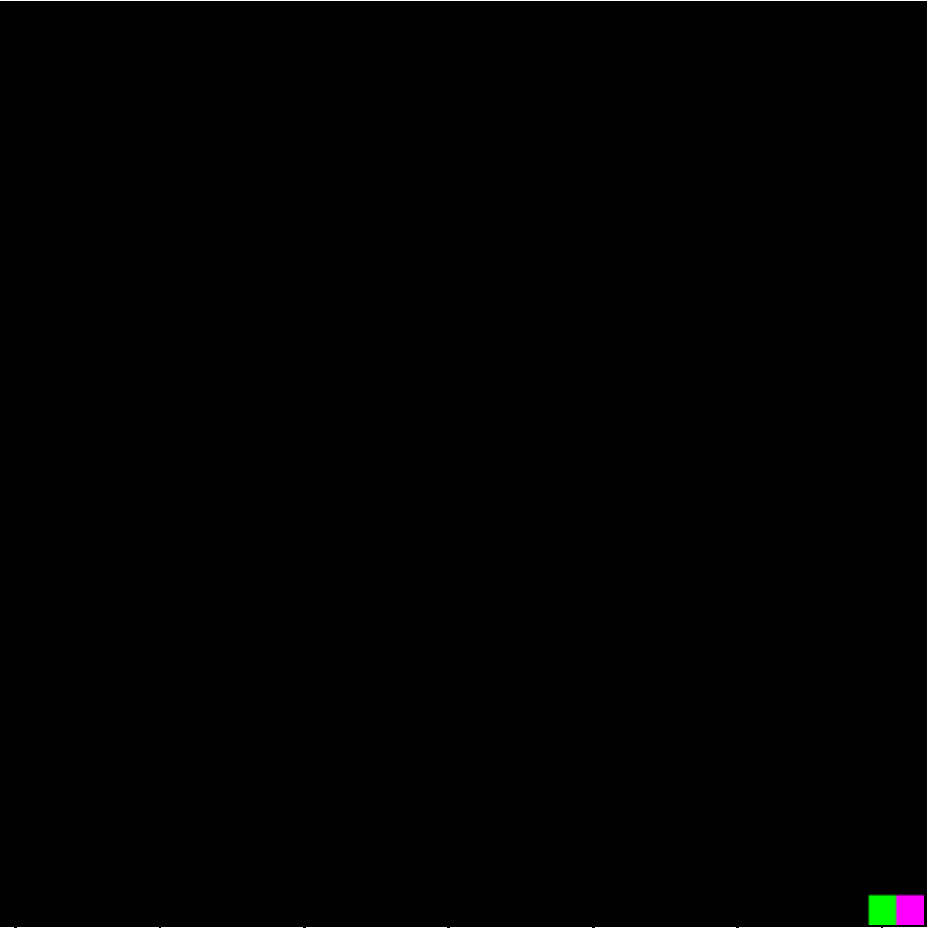} \\
       Original  & $L_0$ Attack & Difference \\[5pt]
      \includegraphics[width=0.28\linewidth]{figures/0_10.png}  &  \includegraphics[width=0.28\linewidth]{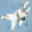} & \includegraphics[width=0.28\linewidth]{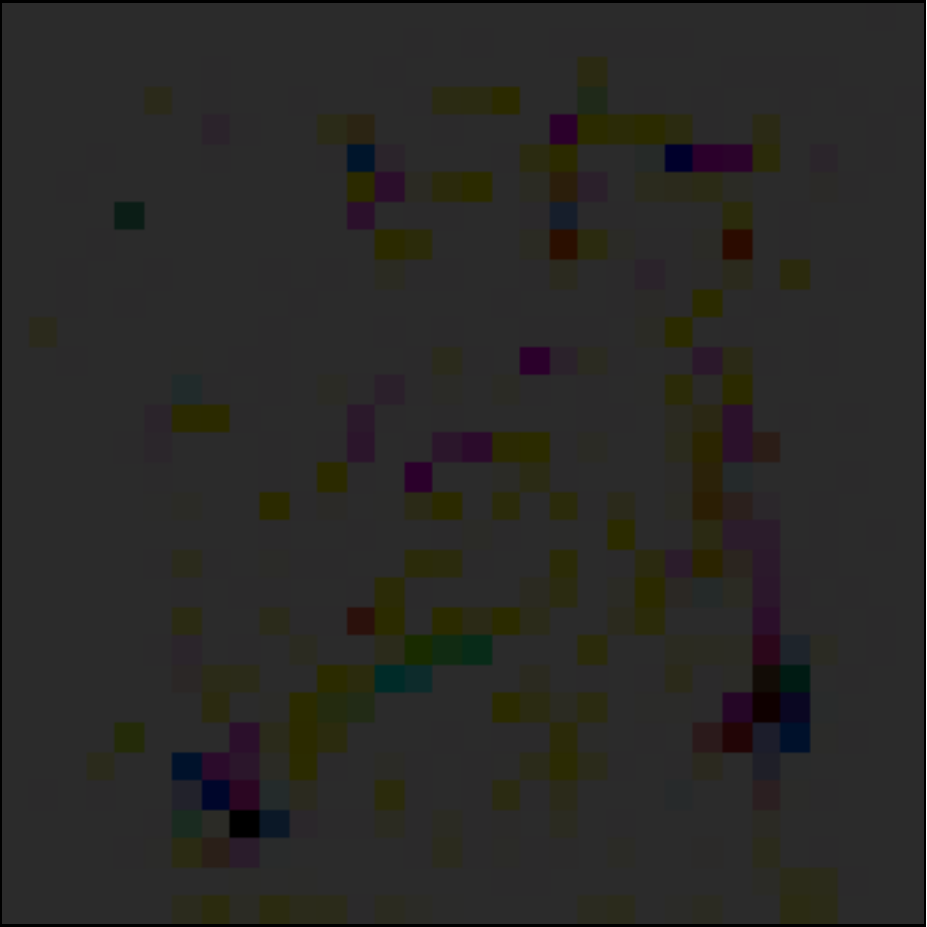} \\
      Original  & $L_{\infty}$ Attack & Difference \\[5pt]
    \end{tabular}
    \caption{The first column is the original image, the second is the $L_p$-attacks ($L_2$-norm=5, $L_0$-norm=2 and $L_{\infty}$-norm=0.16), and the third is the highlighting difference (Trigger).}
    \label{fig_L_p_trigger}
\end{figure}

After generating the final trigger $\alpha^*$, we construct the poisoning image~$x^{\prime}$ by adding the trigger into the image $x$ randomly drawn from the original training set~$\mathcal{D}_{tr}$ with a sampling ratio $\epsilon$, and assign a target label~$t$ determined by the adversary.   
The proposed attack we implemented is universal, meaning we can build our poisoned image~$x^{\prime}$ \textbf{by choosing any image $x$ without considering their original labels}. An example for the poisoning image~$x^{\prime}$ is shown in Fig.~\ref{fig_L_p_trigger}. 

After poisoning the input images according to the above process, we have a set of poisoning images $(x^{\prime},t) \in \mathcal{D}^{p}_{tr}$. Next we combine the original training set and the poisoning training set together into a new training set ($\mathcal{D}_{train} \cup \mathcal{D}_{p}$). We use $\epsilon \in (0, 0.1]$ to control the pollution ratio, defined as the portion of the poisoning training set~$\mathcal{D}^p_{tr}$ over the whole new training set. Finally, we use this new training set to retrain a classifier $h^*$ from the original pre-trained model $h$. We observe a high efficiency in retraining from the pre-trained model $h$ to our expected model $h^*$ using a poisoning training set with a pollution rate of $\epsilon=0.05$, with only $5$ epochs elapsed before model convergence. 
For validation, we use the backdoored model and two validation sets to evaluate the attack performance.

\section{Experimental Analysis}
\begin{figure}[t]
    \centering
	\begin{subfigure}[t]{0.23\textwidth}
		\centering
		\includegraphics[width=1\linewidth]{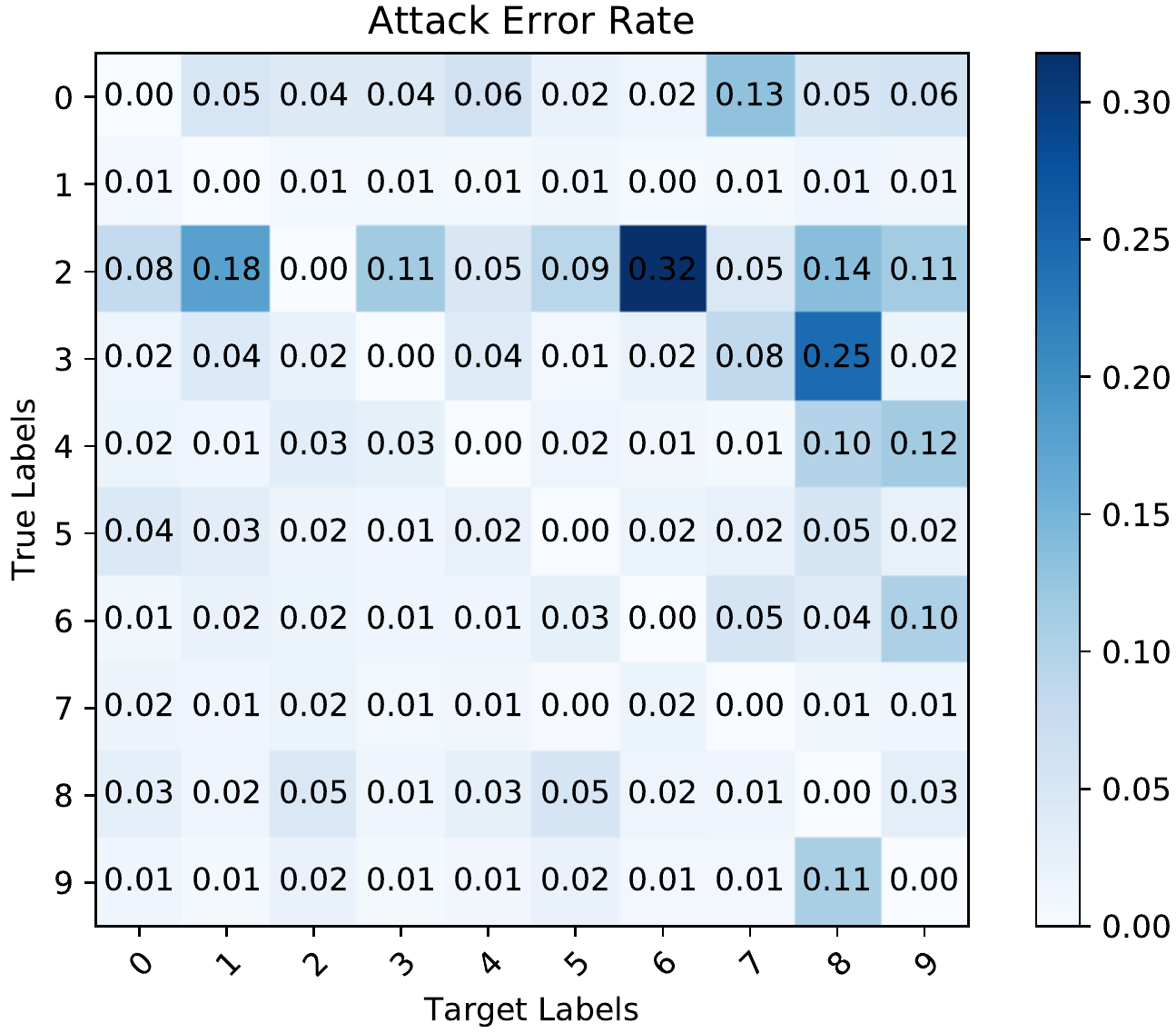}
		\caption{1 - Attack Error Rate}
		\label{fig:LSB_exp-a}
	\end{subfigure}
	\begin{subfigure}[t]{0.23\textwidth}
		\includegraphics[width=1\linewidth]{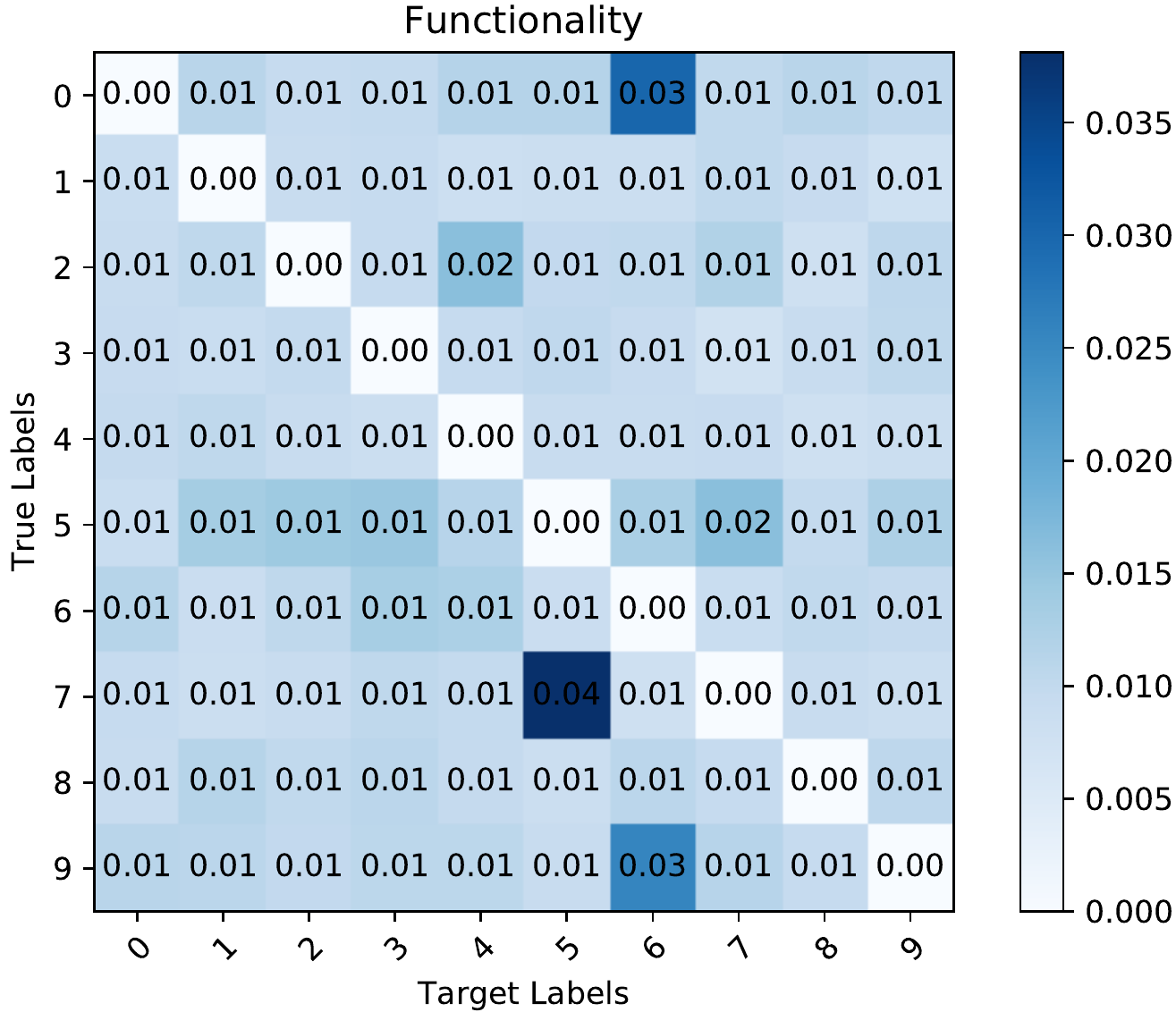}
		\caption{2 - Functionality}
		\label{fig:LSB_exp-b}
	\end{subfigure}
	\vspace{-0.1cm}
	\caption{Classification errors for the clean images (left) and the backdoored images (right). Lower error rates on both are reflective of the attack's success.}
	\label{fig:LSB_exp}
\end{figure}

In this section, we implement our two types of attacks introduced in Sections~\ref{sec:add_stegan} and \ref{sec:add_regular}. For the first type of triggers generated through steganography, we mount our attack on MNIST~\cite{MNIST}, CIFAR-10~\cite{CIFAR10}, and GTSRB~\cite{Houben-IJCNN-2013}. For the second type of trigger optimized through regularization, we mount our attack on CIFAR10/100~\cite{CIFAR10} and GTSRB~\cite{Stallkamp2012}. All four datasets are widely used in image classification. Our experiments were run on a machine with a Intel i9-7900X, with 64GB of memory and a GTX1080; our networks are implemented with Pytorch 1.4.

\subsection{Single Target Backdoor Attacks via Steganography} \label{sec:exp_stego}
\noindent \textbf{Setup.} For single target backdoor attacks, the trigger is only valuable for one source class. For each source-target pair (where the source class must be different from the target class), there is one independent trigger.\footnote{We note that these triggers can be the same; however, in our experiments we vary the trigger.} We implement the attack strategy described in Section~\ref{sec:add_stegan}. We mount our attacks on the MNIST, CIFAR-10, and GTSRB datasets. The MNIST digit recognition task is a 10 label $[0-9]$ classification task. MNIST contains 60,000 training images and 10,000 test images. 
The CIFAR-10 dataset~\cite{CIFAR10} consists of 60,000 32x32 colour images in 10 classes, with 6,000 images for each class; so there are 50,000 training images and 10,000 test images. The German Traffic Sign Recognition Benchmark (GTSRB) contains 43 classes, split into 39,209 training images and 12,630 test images. For MNIST, in order to provide high-quality classification model on this task, we use the network architecture proposed by Zhang et al.~\cite{CNN_Arch} as our basic model, which consists of two convolutional layers and two fully-connected layers. We achieve an accuracy of $99.5\%$ on validation set using this CNN network architecture. For CIFAR-10 and GTSRB datasets, we choose the ResNet-18~\cite{he2016deep} network architecture to obtain baseline models of $92.48\%$ and $95.31\%$ validation accuracy, respectively. The text trigger used for our three datasets is the string ``Apple" repeated 100 times ``AppleAppleA....pple"; the length of this string is 500.

\noindent \textbf{MNIST.} To build the poisoning training set with the LSB algorithm, we embed the trigger into the least significant bit of the input data, achieving the bit-level feature addition into the poisoning training set. For the images in which we embedded the trigger, we modify their labels from $i$ to $j$ $(j \neq i)$. We consider label $j$ as the target label and label $i$ as the source. We use $\epsilon$ to control the pollution rate of the poisoning dataset. For this bit-level feature to be learned by the neural networks, we retrain the pre-trained baseline model on the poisoning dataset with a small batch size and a small learning rate. When validating this model, we hide the same trigger on the original test dataset using the same LSB configuration, and then compute metrics defined in Section~\ref{sec:measurement}. The target labels range from $[0-9]$, and the original labels range from $[0-9]$, with the exception of the target label $j$. This results in 90 label pairs to test. 

For every trained backdoor model across 90 pairs of experimental configurations, we compute their \textit{Functionality}, \textit{Attack Success Rate} metrics (see Section~\ref{sec:measurement}). Fig.~\ref{fig:LSB_exp} illustrates the error rates of the backdoored model on the clean images (\textit{Functionality}) and the validation poisoning set (\textit{Attack Success Rate}). The colour-shaded cells in row $i$ and column $j$ of Fig.~\ref{fig:LSB_exp-a} and Fig.~\ref{fig:LSB_exp-b} represent the error on clean images and poisoned images, respectively. For the poisoned images, we consider the ground truth as the mapped label~$j$. 
A successful attack is observed when the original ground truth label of $i$ is mapped to the target label $j$; thus the error rate reported in Fig.~\ref{fig:LSB_exp} indicates when the poisoned image is not predicted as the the target label. The validation error rate on clean dataset observes a slight increase from the baseline MNIST model, which means the \textit{Functionality} of the normal users experiences a slight degradation; in the baseline MNIST model, the \textit{Functionality} is $99.5\%$, in contrast to the backdoored model; the worst case \textit{Functionality} is $96.19\%$ (in the position with a coordinate (5, 7) in Fig.~\ref{fig:LSB_exp-b}), leading to a $3.31\%$ decrease in the functionality of the backdoored model.

Many of our single target backdoored models have a higher \textit{Attack Success Rate}; however, there are still some instances with a low attack success rate. In the worst performing attack, the attack success rate observed is for the attack in which poisoned images of digit 2 are mislabeled as digit 6; the attack success rate only has $68.22\%$ (the position with a coordinate (6, 2) in Fig.~\ref{fig:LSB_exp-b}). The reason why this instance is significant is that likely a result of the decision bounds on the pre-trained model possessing different distances for each instance. The larger the distance between two class's decision boundaries, the harder it is for the attacker to inject a backdoor by finding a shortcut to transfer from the source class to the target class. These instances will have a low attack success rate; however, we can still obtain a high \textit{Attack Success Rate} by using a larger trigger to mitigate the decision boundary distance. Thus, our attack can be tailored to have a good performance on all (source, target) pairs.

\begin{figure}[t]
    \centering
    \includegraphics[width=0.80\linewidth]{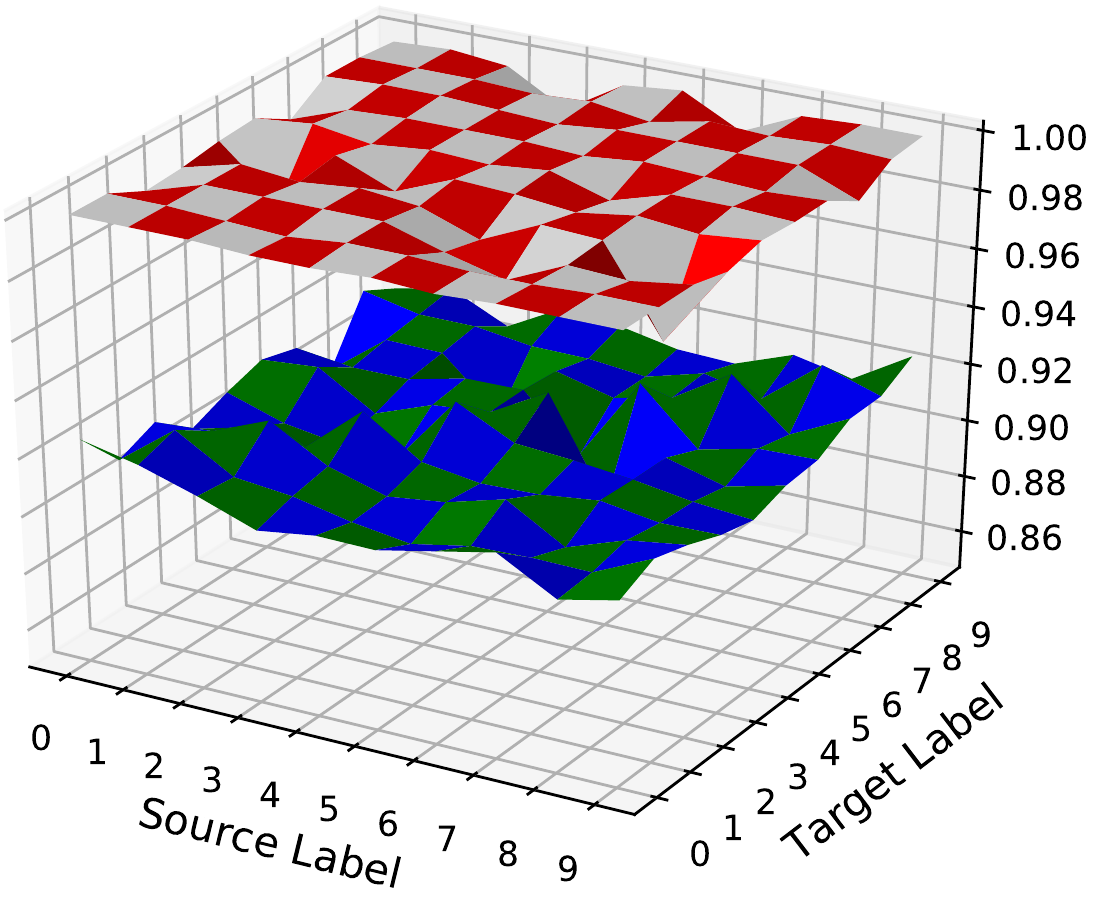}
    \caption{The \textit{Attack Success Rate} (top) and the \textit{Functionality} (bottom) on CIFAR-10.}
    \label{fig:stego_cifar10}
    \vspace{4mm}
\end{figure}

\noindent \textbf{CIFAR10.} We adopt the single target attack on the CIFAR-10 dataset with the same configuration as the MNIST dataset. As the CIFAR-10 dataset also has 10 classes, there are 90 (source, target) pairs. Figure~\ref{fig:stego_cifar10} reports the \textit{Attack Success Rate} (top surface) and \textit{Functionality} (bottom surface) metrics of the attack on the CIFAR-10 dataset. The X-axis and Y-axis present source and target labels, respectively, which means if we add the trigger to images drawn from the source labeled class, the corresponding backdoored model will output the target label successfully (the top surface on Z-axis). While for input images without applying the trigger, the predictions of the corresponding backdoored model should be the same as their original labels (the bottom surface on Z-axis).  
Overall we find that all of these 90 (source, target) pairs have a high \textit{Attack Success Rate} and \textit{Functionality} on the CIFAR-10 dataset. The \textit{Attack Success Rate} surface over the \textit{Functionality} surface shows that the DNNs can identify the trigger as a well-detectable feature other than the normal features of the input images.

\noindent \textbf{GTSRB.} In contrast to the MNIST and CIFAR datasets, GTSRB is a dataset with a direct real-world scenario application, i.e., classifying traffic signs. Images in this dataset are instances of physical traffic signs, with each real-world traffic sign occurring only once. In this test, we choose three sub-classes as an example: stop signs, speed-limit signs, and warning signs. We experimented with a backdoor trigger, which is a text string with a length of 500, against the outsourced baseline model. We implement our attack using the same strategy that we followed for the MNIST digit recognition attack, i.e., by poisoning the training dataset via steganography and changing corresponding ground truth labels. As a control group, we do not poison the training dataset, just change corresponding ground truth labels. Table~\ref{tab:stego_gtsrb} reports these three instances according to the \textit{Functionality} (left) and the \textit{Attack Success Rate} (right) metrics.
\begin{table}[t]\footnotesize
	\centering
	\caption{Single Target Backdoor Attacks on GTSRB}
	\label{tab:stego_gtsrb}
	\vspace{-1mm}
	\begin{tabular}{l l r r}
		\toprule
		\bf Source Class & \bf Target Class & \bf Backdoored & \bf Clean \\
		\midrule
		 stop  & speedlimit & 0.9678/0.9259 & 0.9732/0.0000 \\ \hline
		 warning  & stop & 0.9726/0.9889 & 0.9753/0.0205\\ \hline
		 speedlimit & warning & 0.9683/0.9641 & 0.9702/0.0000 \\ \bottomrule
	\end{tabular}
\end{table}
As a control group, we poison the selected images drawn from the original training set without embedding anything, just change their labels to the target label. The result shows in  the last column of Table~\ref{tab:stego_gtsrb}, which indicates that the small perturbation caused by steganography can be identified by the DNNs while not for humans. We can use this type of irregular trigger patterns to perform the backdoor attacks.

\noindent \textbf{Comparison with BadNets.} BadNets use two types of triggers, the single pixel attack and the pattern attack, to infect the clean images. A list of triggers used by BadNets can be found in Table~\ref{tab:cmp_badnets}. Note that the single pixel attack of BadNets creates the most difficult task for a DNN to identify the difference between the clean and the poisoned samples. In BadNets, they also admit that they had to change the training parameters, including the step size and the mini-batch size to get the training error to converge. We conduct our experiment on the MNIST dataset, to corroborate the result that the single pixel attack of BadNets needs the most epochs to converge. We do note that for the single target attack, the pollution rates are computed on one class, which differs from universal attacks which are computed from the entire training set. 

Compared with the single pixel backdoor attack of BadNets, it is easier for our steganography based attack to achieve convergence of the training error. For the single pixel attack of BadNets, it requires 80 epochs and a 0.5 pollution rate to achieve an acceptable attack success rate. While for our steganography approach, it is only 20 epochs and a 0.1 pollution rate. Besides, we can control the effectiveness of the backdoor attack by increasing the size of the text trigger (as more trigger bits will be encoded). When the size of the text trigger increases, even less epochs and a lower pollution rate are needed to achieve training error convergence. When compared with the pattern backdoor of BadNets, our steganography attack achieves a higher PASS score and a lower LPIPS score, and is thus more invisible. Table~\ref{tab:cmp_badnets} and Table~\ref{table:1} show that our attacks achieve a commendable performance in comparison to BadNets, with a less perceivable mask and a comparable attack success rate.
\begin{table*}[t]
\centering
\caption{Performance in comparison to BadNets}
\label{tab:cmp_badnets}

\resizebox{0.9\textwidth}{!}{
\begin{tabular}{llrrrrrcc}  
\toprule
\multirow{2}{*}{\bf Attacks}  & \multirow{2}{*}{\bf Trigger} & \multirow{2}{*}{\bf Trigger size} & \multirow{2}{*}{\bf \tabincell{c}{Source /\\ Target Label}} & \multirow{2}{*}{\bf Pollution Rate} & \multirow{2}{*}{\bf \tabincell{c}{Clean Model\\ Accuracy (\%)}} & \multirow{2}{*}{\bf Epoch} & \multicolumn{2}{c}{\bf Performance (\%)} \\
\cmidrule(r){8-9}
 & & & & & & & {\bf Functionality(\%)} &  {\bf Attack Success Rate(\%)} \\
\midrule
BadNets & Single Pixel & $L_0$=1 & 4/7 & 0.5 & 99.5 & 80 & 99.11 & 99.49\\
BadNets    & Pattern Trigger & $L_0$=4 & 4/7 & 0.1 & 99.5 & 5 & 99.22 & 99.19 \\
Steganography & Text:``Apple...``  & 500 & 4/7 & 0.1 & 99.5 & 20 & 99.01 & 98.17\\
\bottomrule
\end{tabular}
}
\end{table*}

\begin{figure}[t]
	\begin{subfigure}[t]{0.24\textwidth}
		\centering
		\includegraphics[width=1\linewidth]{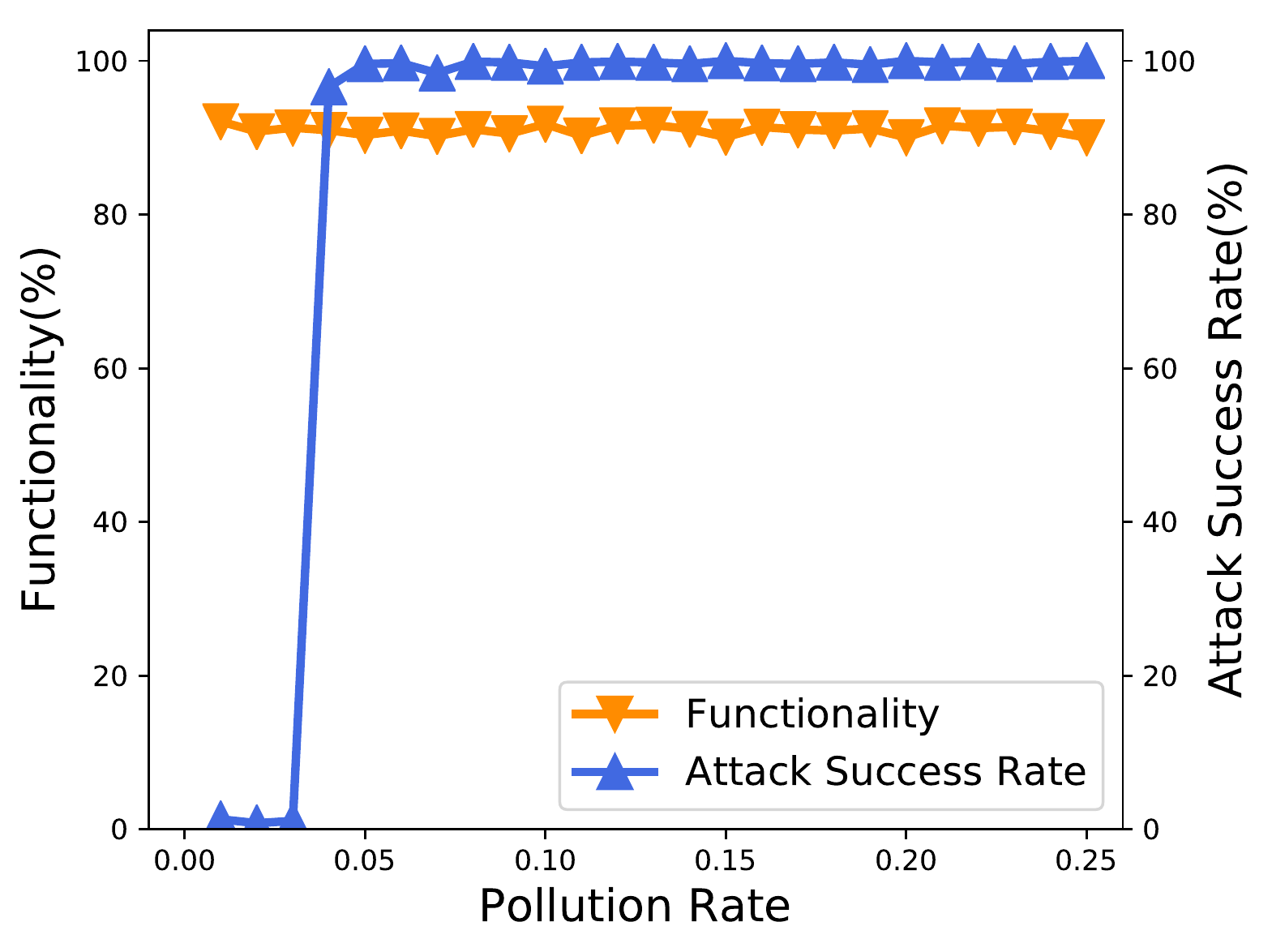}
		\caption{CIFAR-10}
		\label{fig:ste_prt_cf}
	\end{subfigure}
	\begin{subfigure}[t]{0.24\textwidth}
		\includegraphics[width=1\linewidth]{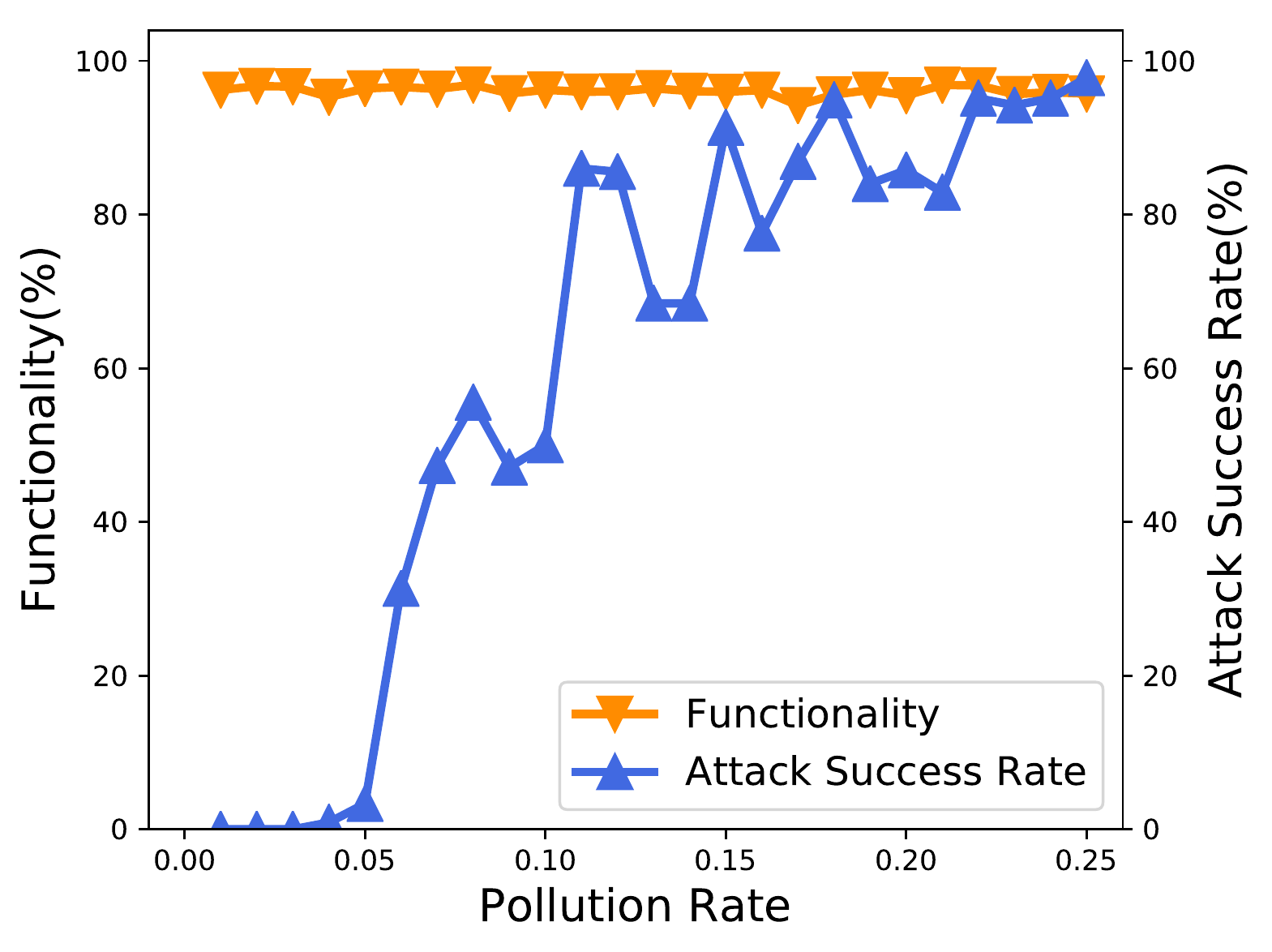}
		\caption{GTSRB}
		\label{fig:ste_prt_gt}
	\end{subfigure}
	\vspace{-0.2cm}
	\caption{The relationship of \textit{Attack Success Rate} and \textit{Functionality} with the pollution rate $\epsilon$ increase in terms of the single target attack on CIFAR-10 (left) and GTSRB (right).}
	\label{fig:prt_single}
\end{figure}

\noindent \textbf{Pollution Rate.} We further investigate the impact of the pollution rate~$\epsilon$ upon the performance of the single target backdoor attack. In the single target attack, the pollution rate~$\epsilon$ is the number of samples drawn from the single source class. Those samples will be poisoned by steganography. The impact of the pollution rate on the \textit{Attack Success Rate} and \textit{Functionality} on CIFAR-10 and GTSRB datasets is shown as Fig.~\ref{fig:prt_single}. 
\begin{figure*}[t]
	\begin{subfigure}[t]{0.48\textwidth}
		\centering
		\begin{subfigure}[t]{0.48\textwidth}
		    \centering
		    \includegraphics[width=1\linewidth]{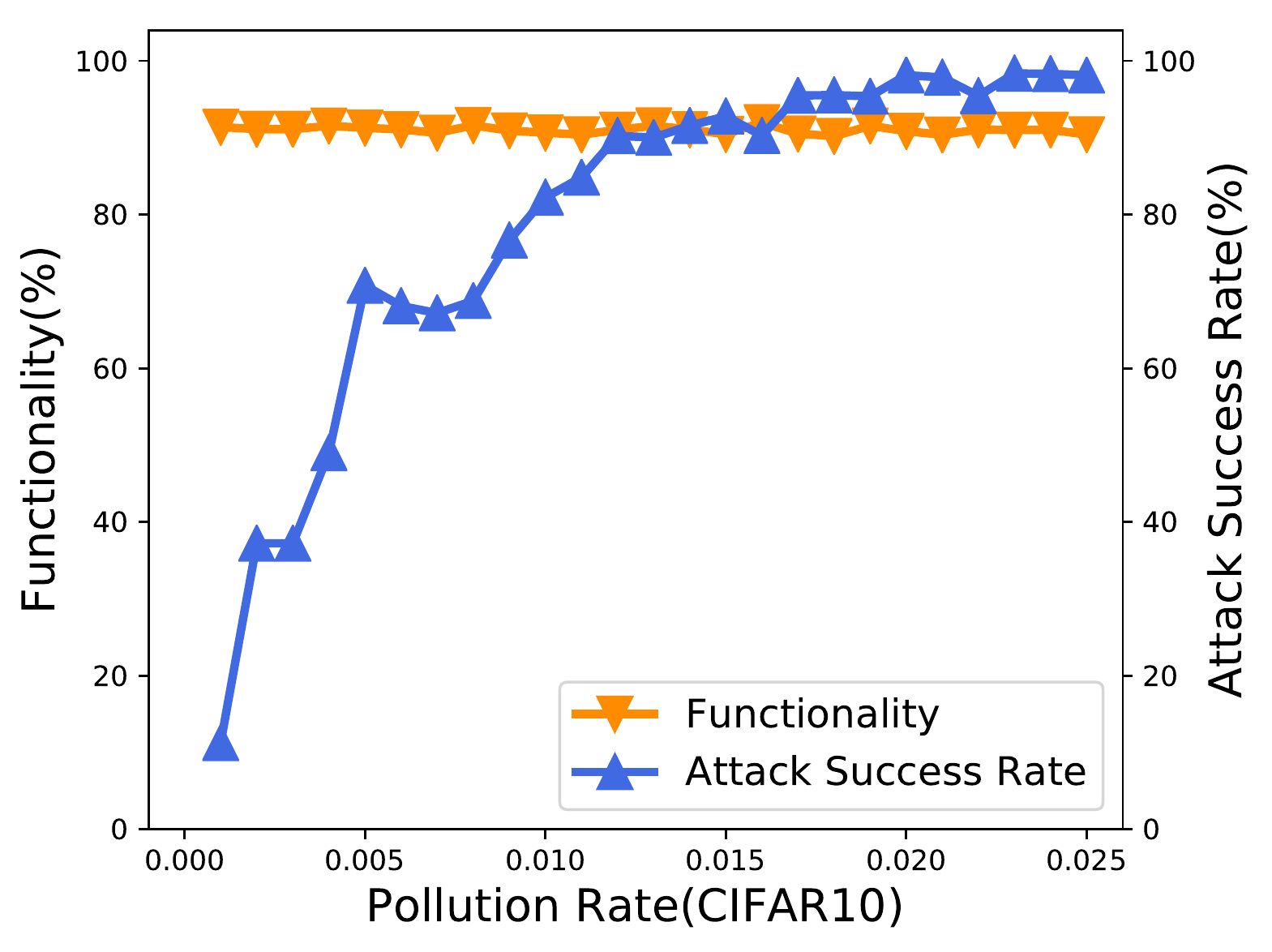}
		\end{subfigure}
    	\begin{subfigure}[t]{0.48\textwidth}
    	    \centering
    		\includegraphics[width=1\linewidth]{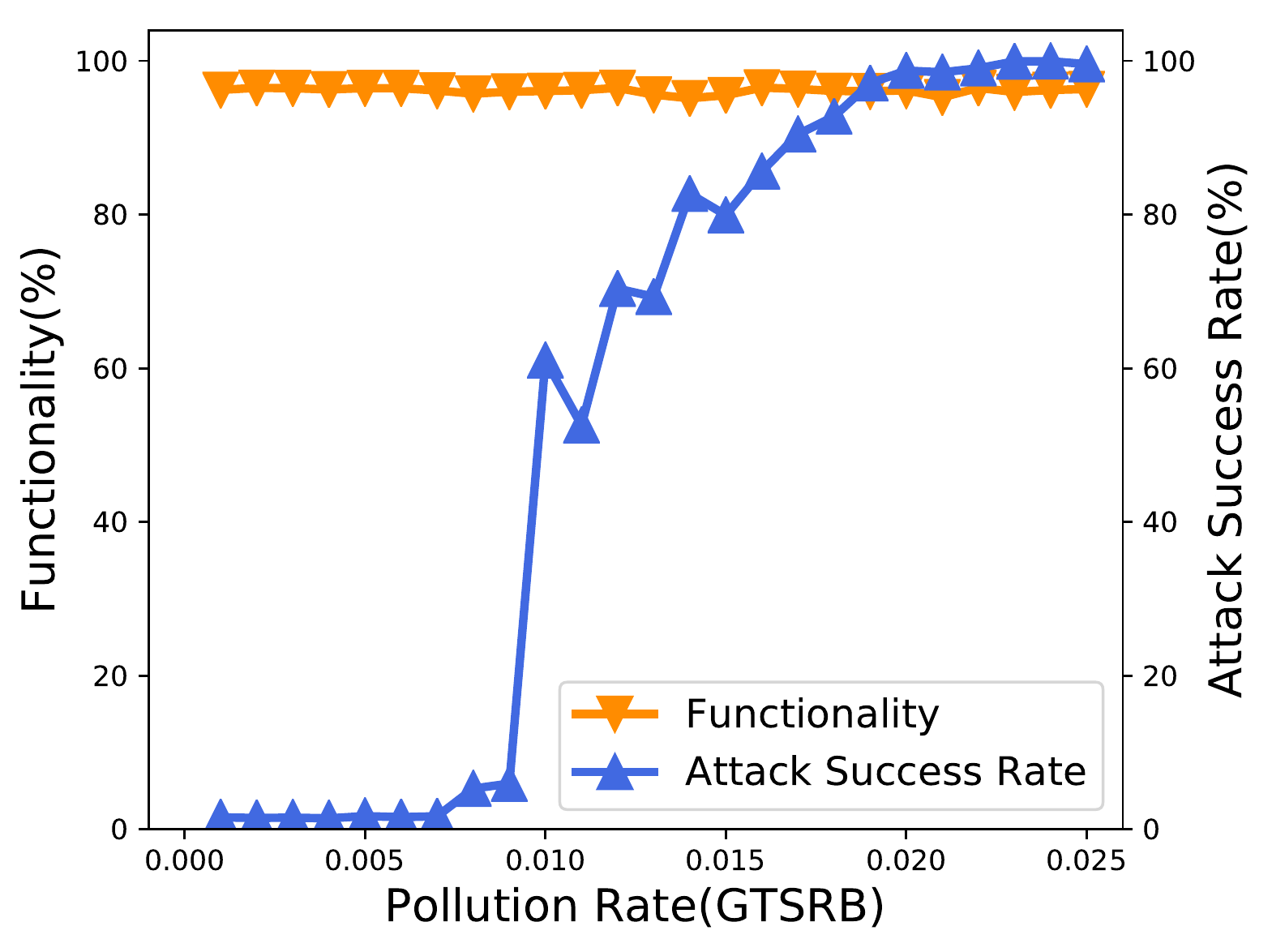}
    	\end{subfigure}
		\caption{$L_2$ Attack}
		\label{fig:l2_prt_cf}
	\end{subfigure}
	\begin{subfigure}[t]{0.48\textwidth}
		\centering
		\begin{subfigure}[t]{0.48\textwidth}
		    \centering
		    \includegraphics[width=1\linewidth]{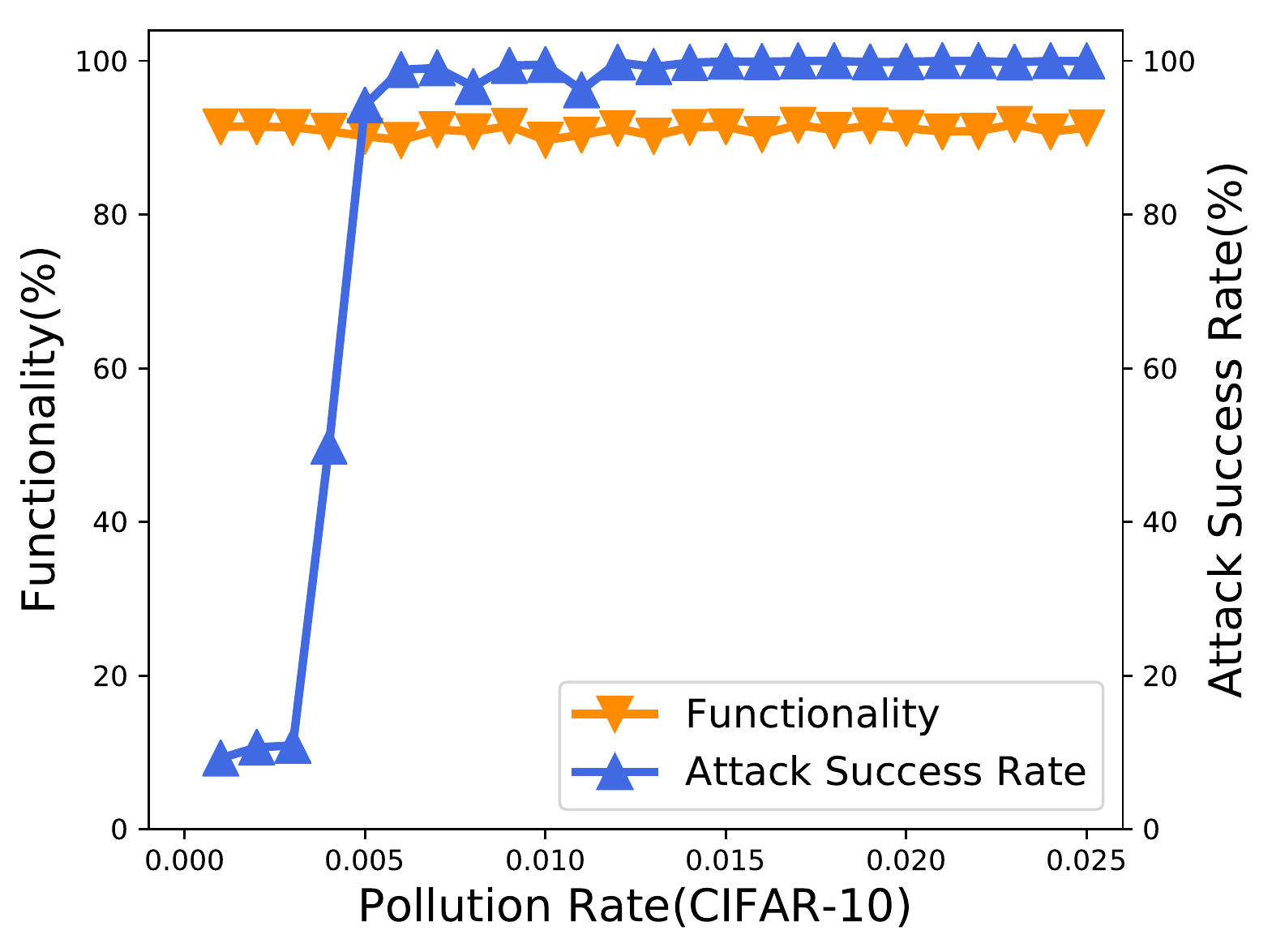}
		\end{subfigure}
    	\begin{subfigure}[t]{0.48\textwidth}
    	    \centering
    		\includegraphics[width=1\linewidth]{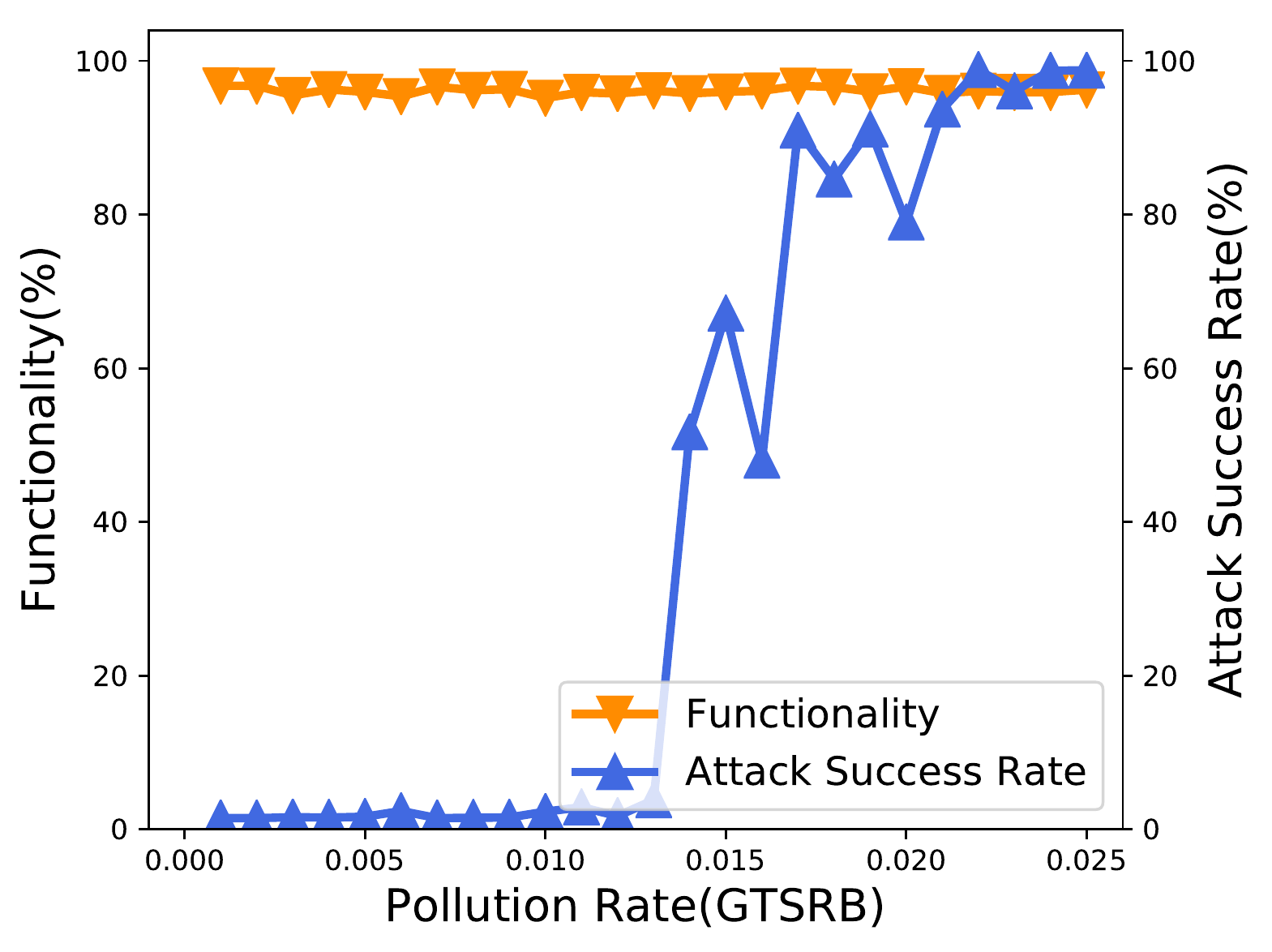}
    	\end{subfigure}		
		\caption{$L_0$ Attack}
	\end{subfigure}
	\vspace{-0.2cm}
	\caption{The relationship of \textit{Attack Success Rate} and \textit{Functionality} with the pollution rate $\epsilon$ increase in terms of the $L_0$ attack and the $L_2$ attack on CIFAR-10 (left) and GTSRB (right).}
	\label{fig:unv_prt}
\end{figure*}
In Fig.~\ref{fig:prt_single}, with an increasing pollution rate the \textit{Attack Success Rate} increases (the blue line in Fig.~\ref{fig:prt_single}), while the \textit{Functionality} remains stable. Besides, for different datasets, the minimal pollution rate to achieve high \textit{Attack Success Rate} is different. For the CIFAR-10 dataset, the minimal pollution rate is $0.04$, leading to an \textit{Attack Success Rate} of $96.6\%$, while for the GTSRB dataset is $0.22$, leading to an \textit{Attack Success Rate} of $95.11\%$.

\noindent \textbf{Invisibility Metrics.} When comparing the \textit{Invisibility} metric with previous BadNets backdoor attacks, we first compute the PASS and LPIPS indices of the two types of triggers used in the previous BadNets backdoor attack (Single Pixel and Pattern Pixel). 
The range of the PASS score is $(0,1]$; if two images are identical, the value is $1$. A larger PASS value indicates that an image will appear more similar to human perception. 
Recall that the LPIPS score measures the perceptual distance between the reference image and the blurred image. This LPIPS score lies between $[0,1)$; if two images are identical, the value is $0$. A lower LPIPS value means two images are more similar; a higher score means the images are more different. 
A comparison of the PASS and LPIPS scores for each attack is found in Table~\ref{table:1}. Our trigger achieves the highest average PASS score (extremely close to $1$) and lowest average LPIPS score (near $0$), better than the triggers of BadNets.  This indicates that humans will have more difficulty in  discerning differences between our trigger and the original image.
\begin{table}[t]
    \small 
    \centering
    \caption{PASS and LPIPS scores compared to BadNets}
    \vspace{-1mm}
    \label{table:1}
    \resizebox{0.85\linewidth}{!}{
    \begin{tabular}{@{}ccccc@{}}
        \toprule
         & Original & \tabincell{c}{Single \\ Pixel} & \tabincell{c}{Pattern \\ Pixel} & \tabincell{c}{Our trigger \\ size 500}  \\
        \midrule
        \tabincell{c}{MNIST \\ \quad } & \includegraphics{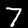} & \includegraphics{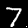} & \includegraphics{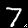} & \includegraphics{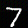} \\
        Avg. PASS & 1 & 0.9890 & 0.9610 & 0.9994 \\
        Avg. LPIPS & 0.0 & 0.0068 & 0.0247 & 2.7e-05 \\
        \midrule
        \tabincell{c}{CIFAR-10 \\ \quad } & \includegraphics{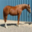} & \includegraphics{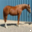} & \includegraphics{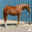} & \includegraphics{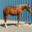} \\
        Avg. PASS & 1 & 0.9916 & 0.9714 & 0.9997 \\
        Avg. LPIPS & 0.0 & 0.0079 & 0.0238 & 1.4e-4 \\
        \midrule
        \tabincell{c}{GTSRB \\ \quad } & \includegraphics{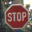} & \includegraphics{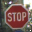} & \includegraphics{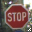} & \includegraphics{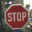} \\
        Avg. PASS & 1 & 0.9909 & 0.9695 & 0.9997 \\
        Avg. LPIPS & 0.0 & 0.0063 & 0.0214 & 1.2e-4 \\
        \bottomrule
    \end{tabular}}
    \vspace{5mm}
\end{table} 

\subsection{Universal Backdoor Attacks via Regularization}
\noindent \textbf{Setup.} We implement the attacks introduced in Section~\ref{sec:L2_attack}. For the three types of trigger optimizations through $L_2$, $L_0$ and $L_{\infty}$ regularization, we mount our attacks on the CIFAR-10/100 and GTSRB~\cite{Stallkamp2012} datasets. We use the pre-trained ResNet-18~\cite{he2016deep} model as the basis of our attacks. The CIFAR-10 dataset~\cite{CIFAR10} consists of 60,000 32x32 colour images in 10 classes, with 6,000 images for each class; so there are 50,000 training images and 10,000 test images. CIFAR-100~\cite{CIFAR10} is just like the CIFAR-10, except it has 100 classes and 10 times fewer images. GTSRB was introduced previously (see Section~\ref{sec:exp_stego}). We achieve $92.48\%$, $73.44\%$, and $95.31\%$ prediction accuracy on the respective validation dataset. 

\noindent \textbf{Performance.} We measure the performance of three types of attackers ($L_2$, $L_0$ and $L_{\infty}$) by computing the \textit{Attack Success Rate} and the \textit{Functionality} on our three datasets.
\begin{figure*}[t]
    \begin{subfigure}[t]{0.32\textwidth}
        \centering
        \includegraphics[width=0.95\linewidth]{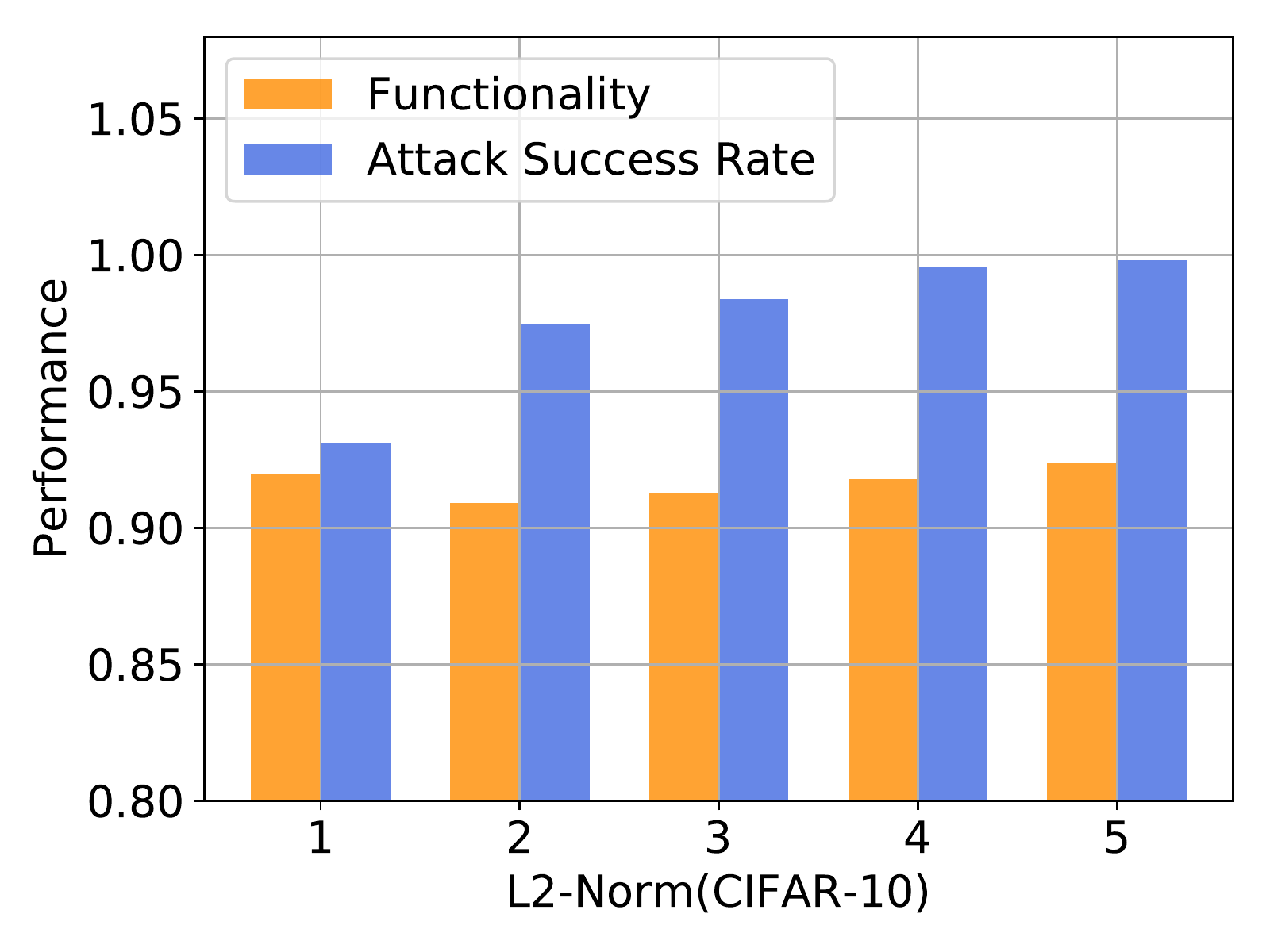}
        \vspace{-0.2cm}
        \label{fig:L2_CIFAR10}
    \end{subfigure}
    \begin{subfigure}[t]{0.32\textwidth}
        \centering
        \includegraphics[width=0.95\linewidth]{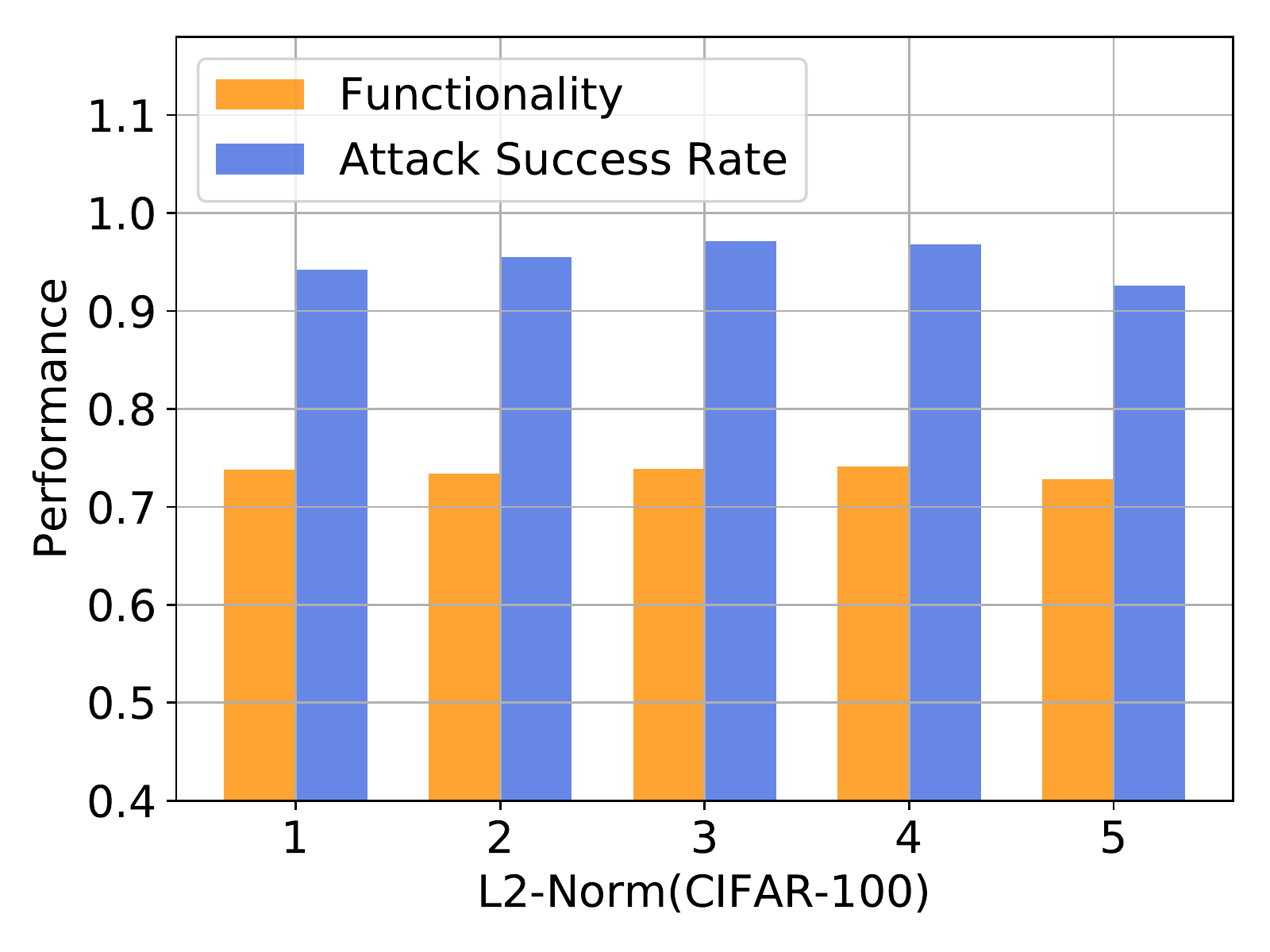}
        \vspace{-0.2cm}
        \label{fig:L2_CIFAR100}
    \end{subfigure}
    \begin{subfigure}[t]{0.32\textwidth}
        \centering
        \includegraphics[width=0.95\linewidth]{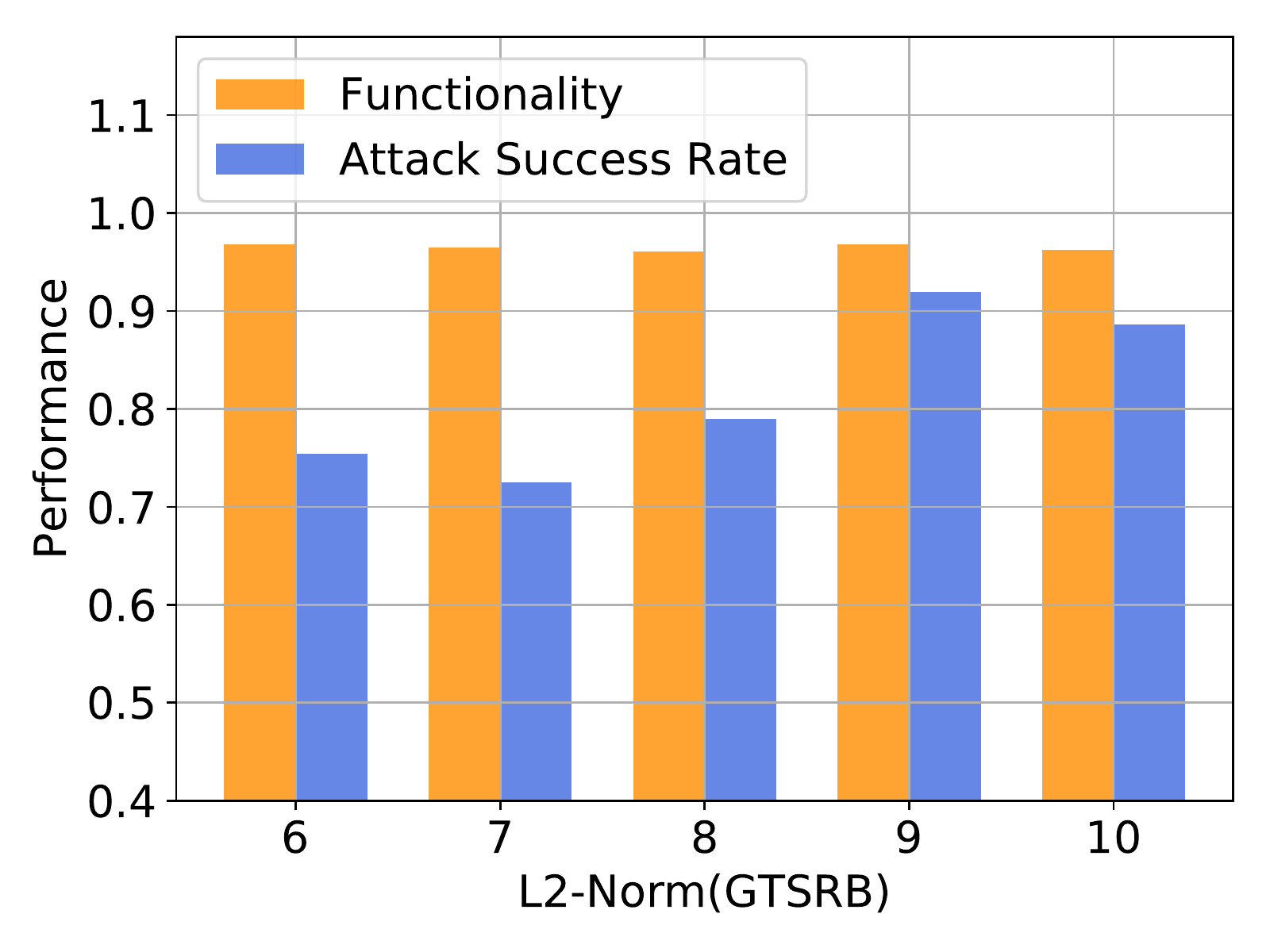}
        \vspace{-0.2cm}
        \label{fig:L2_GT}
    \end{subfigure}
    \vspace{-0.2cm}
    \caption{Functionality and Attack Success Rates of $L_2$ attack on CIFAR-10/100, GTSRB datasets (Validation accuracy: 92.48\%, $73.44\%$ and $95.31\%$, respectively).}
    \label{fig:l2_3ds}
\end{figure*}
\begin{figure*}[t]
    \begin{subfigure}[t]{0.32\textwidth}
        \centering
        \includegraphics[width=0.95\linewidth]{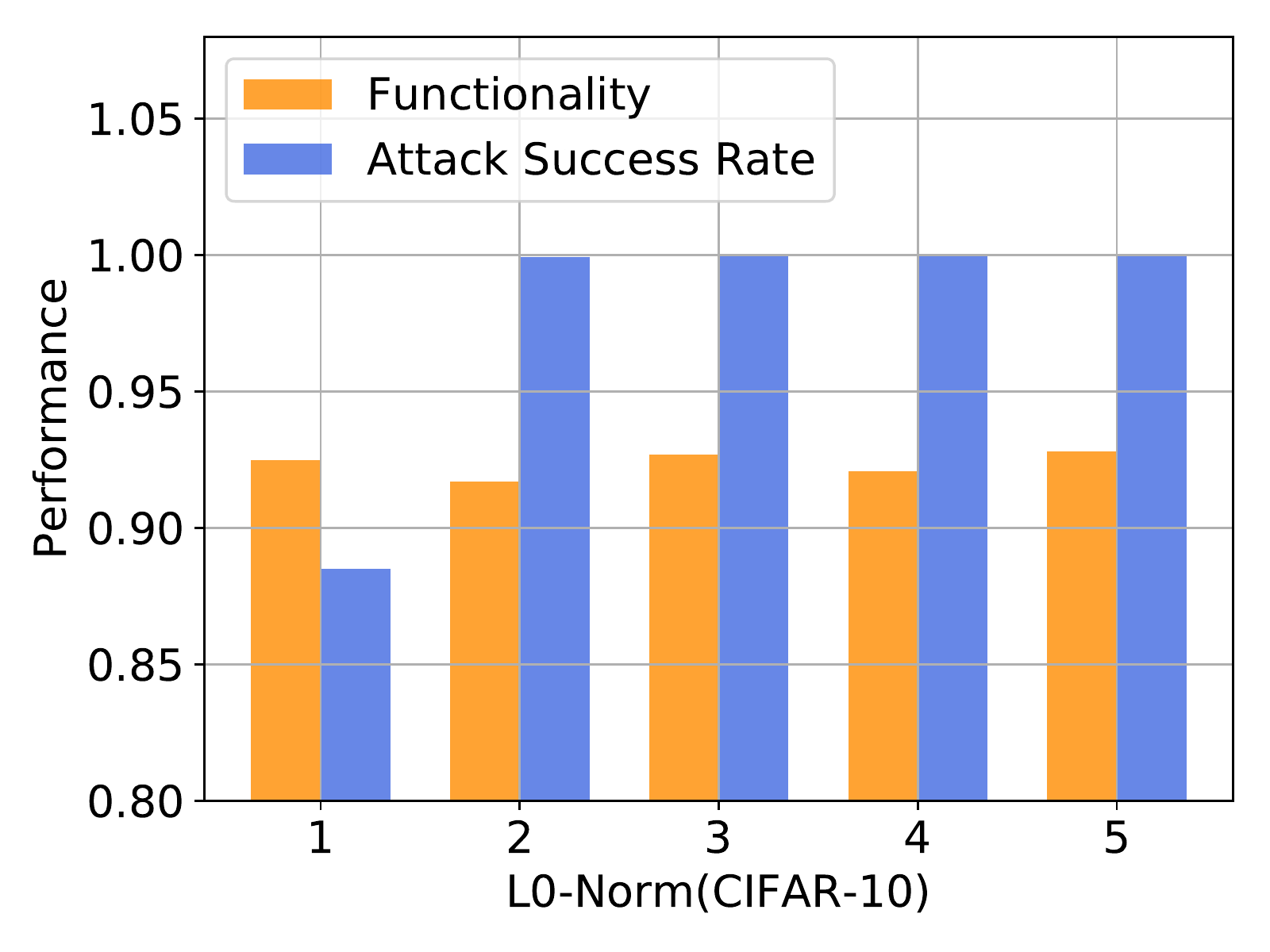}
        \vspace{-0.2cm}
        \label{fig:L0_CIFAR10}
    \end{subfigure}
    \begin{subfigure}[t]{0.32\textwidth}
        \centering
        \includegraphics[width=0.95\linewidth]{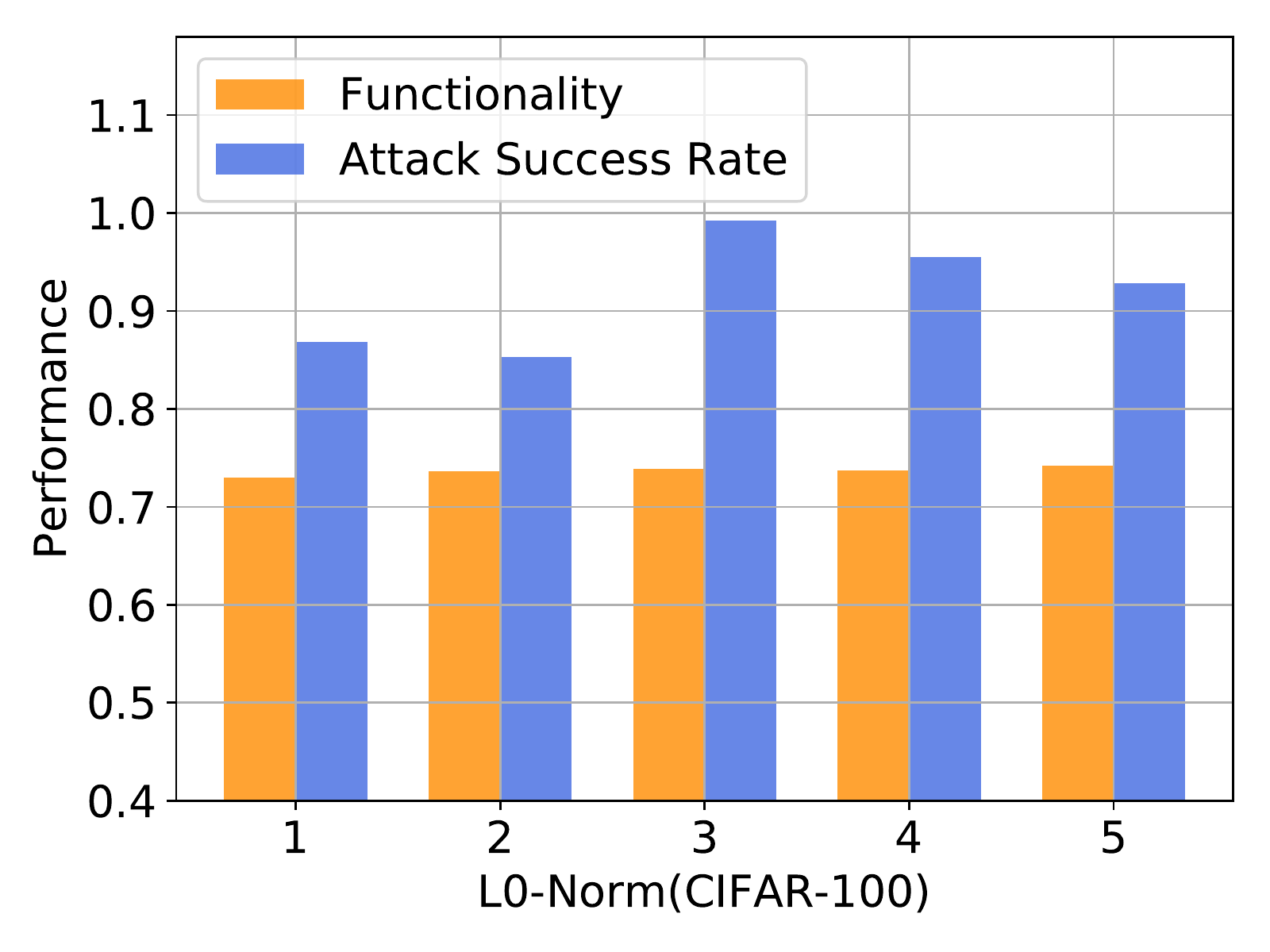}
        \vspace{-0.2cm}
        \label{fig:L0_CIFAR100}
    \end{subfigure}
    \begin{subfigure}[t]{0.32\textwidth}
        \centering
        \includegraphics[width=0.95\linewidth]{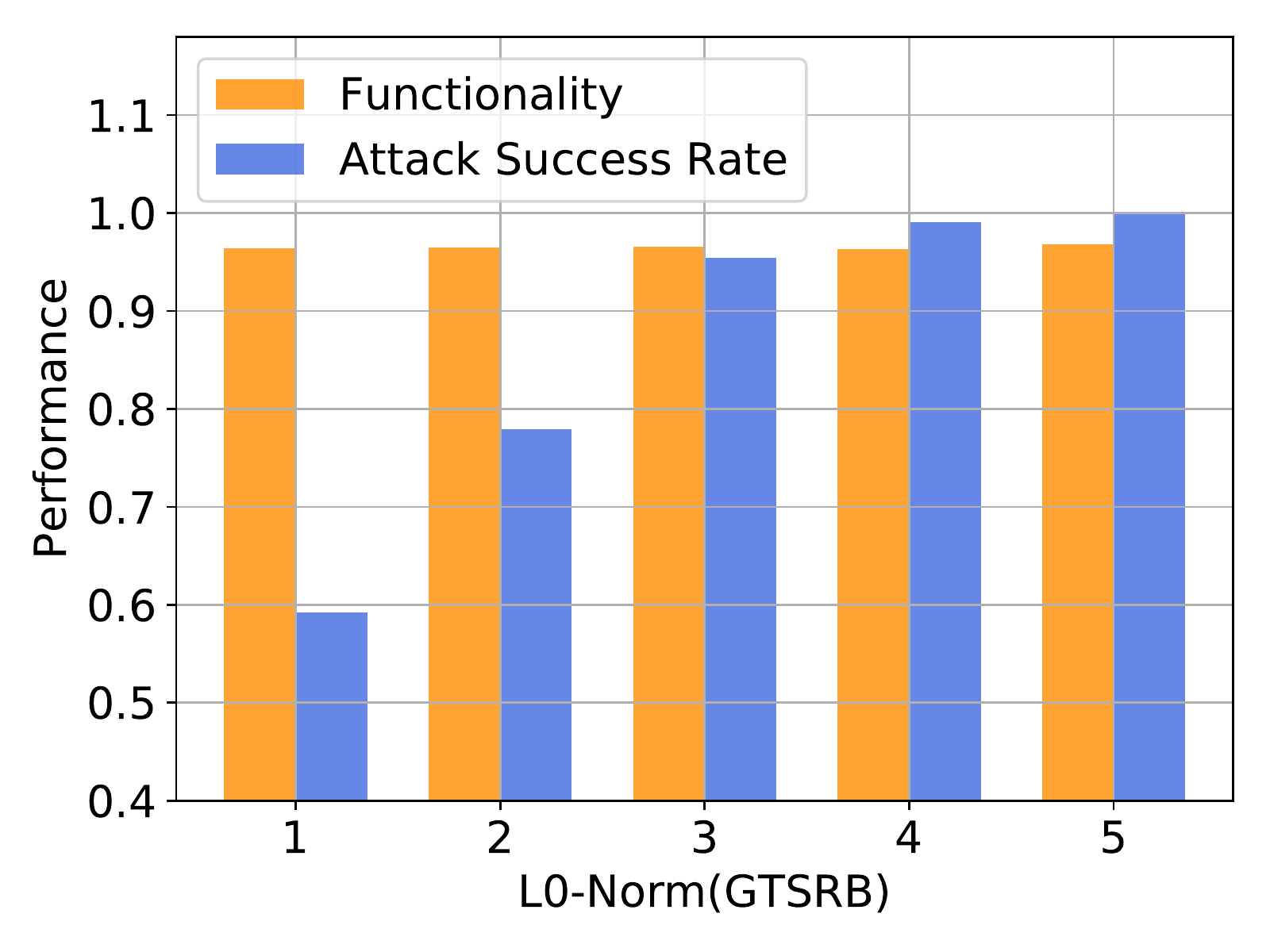}
        \vspace{-0.2cm}
        \label{fig:L0_GT}
    \end{subfigure}
    \vspace{-0.2cm}
    \caption{Functionality and Attack Success Rates of $L_0$ attack on CIFAR-10/100, GTSRB datasets.}
    \label{fig:l0_3ds}
\end{figure*}
\begin{figure*}[t]
    \begin{subfigure}[t]{0.32\textwidth}
        \centering
        \includegraphics[width=0.95\linewidth]{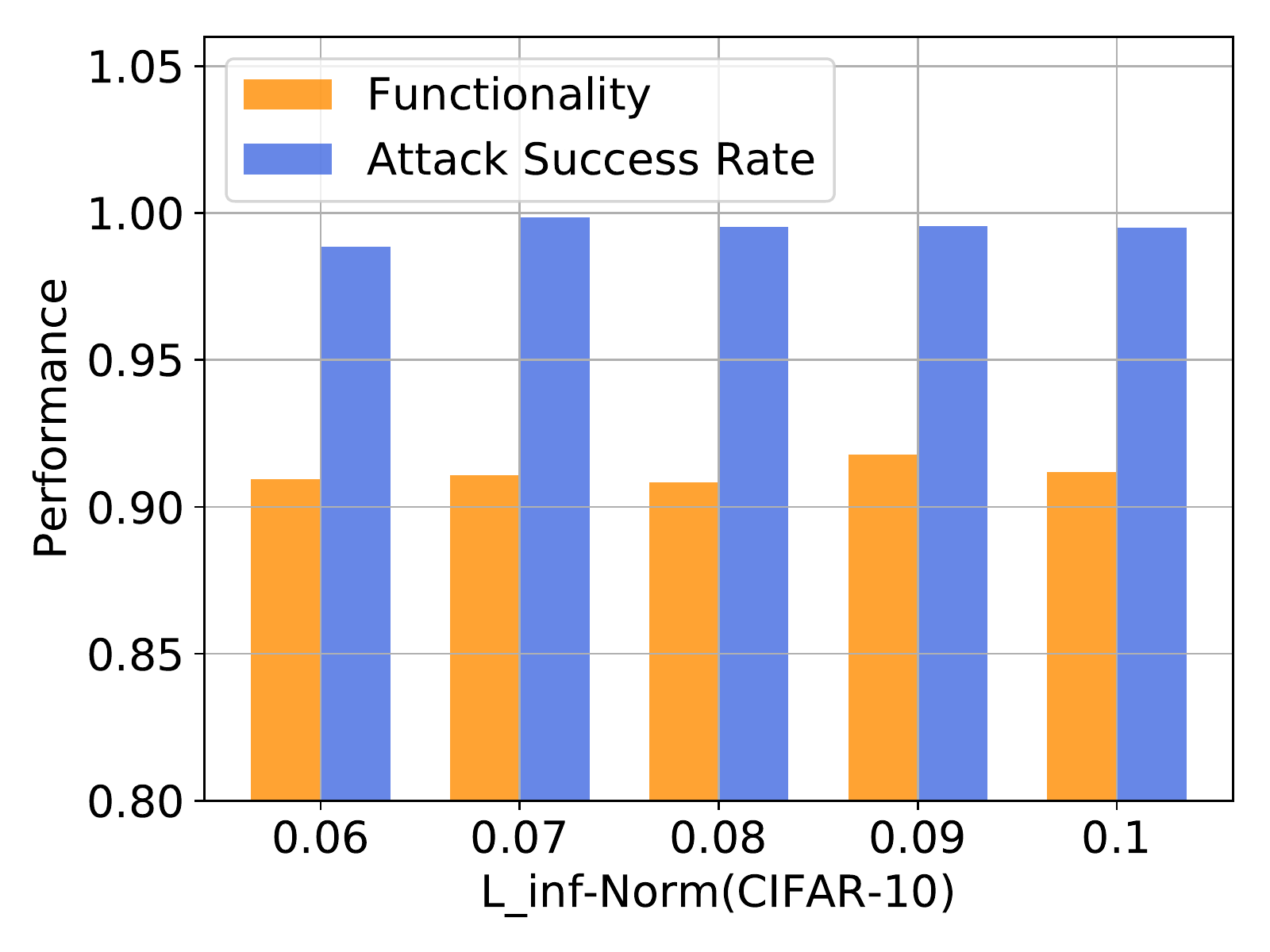}
        \vspace{-0.2cm}
        \label{fig:Li_CIFAR10}
    \end{subfigure}
    \begin{subfigure}[t]{0.32\textwidth}
        \centering
        \includegraphics[width=0.95\linewidth]{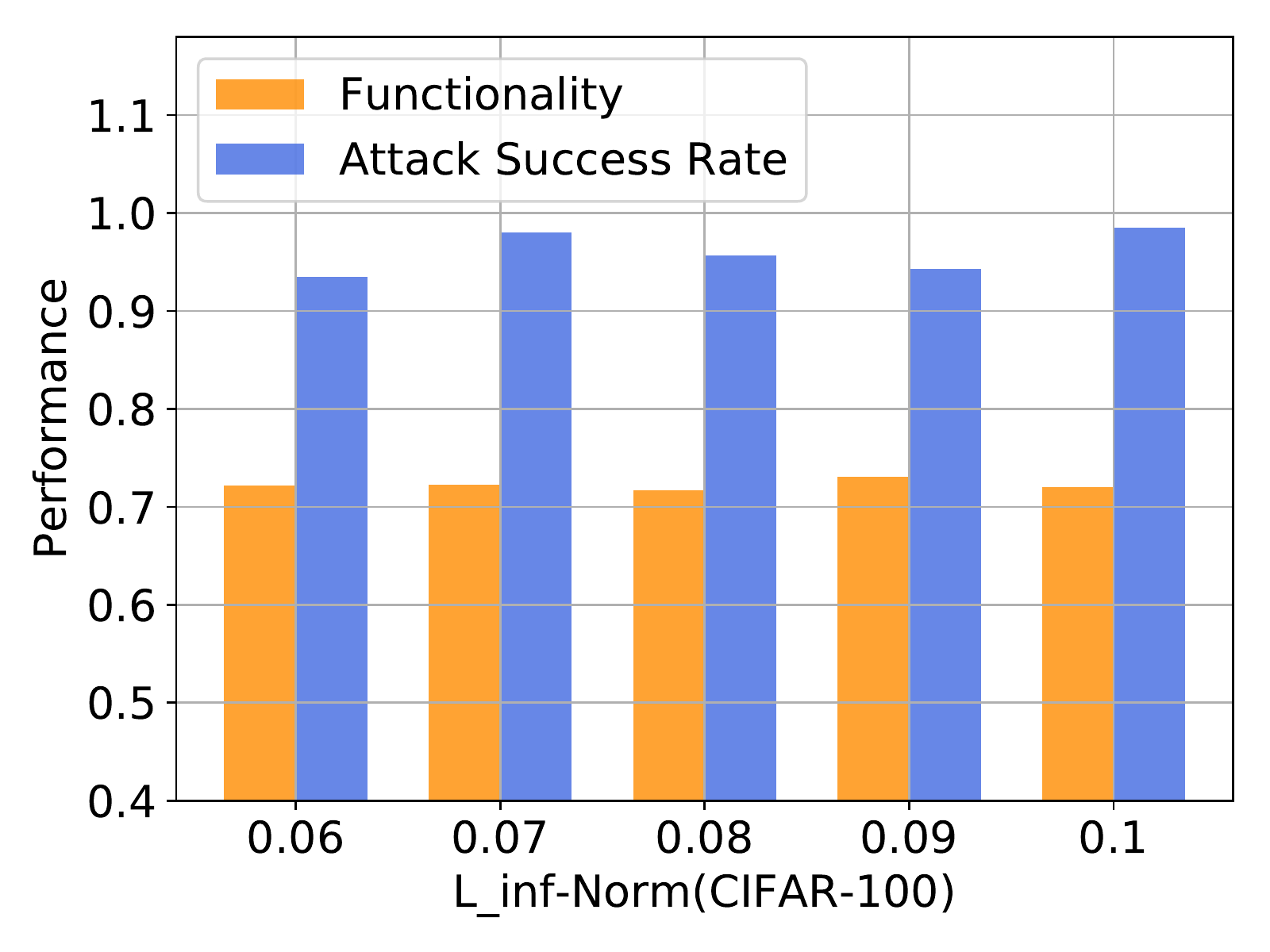}
        \vspace{-0.2cm}
        \label{fig:Li_CIFAR100}
    \end{subfigure}
    \begin{subfigure}[t]{0.32\textwidth}
        \centering
        \includegraphics[width=0.95\linewidth]{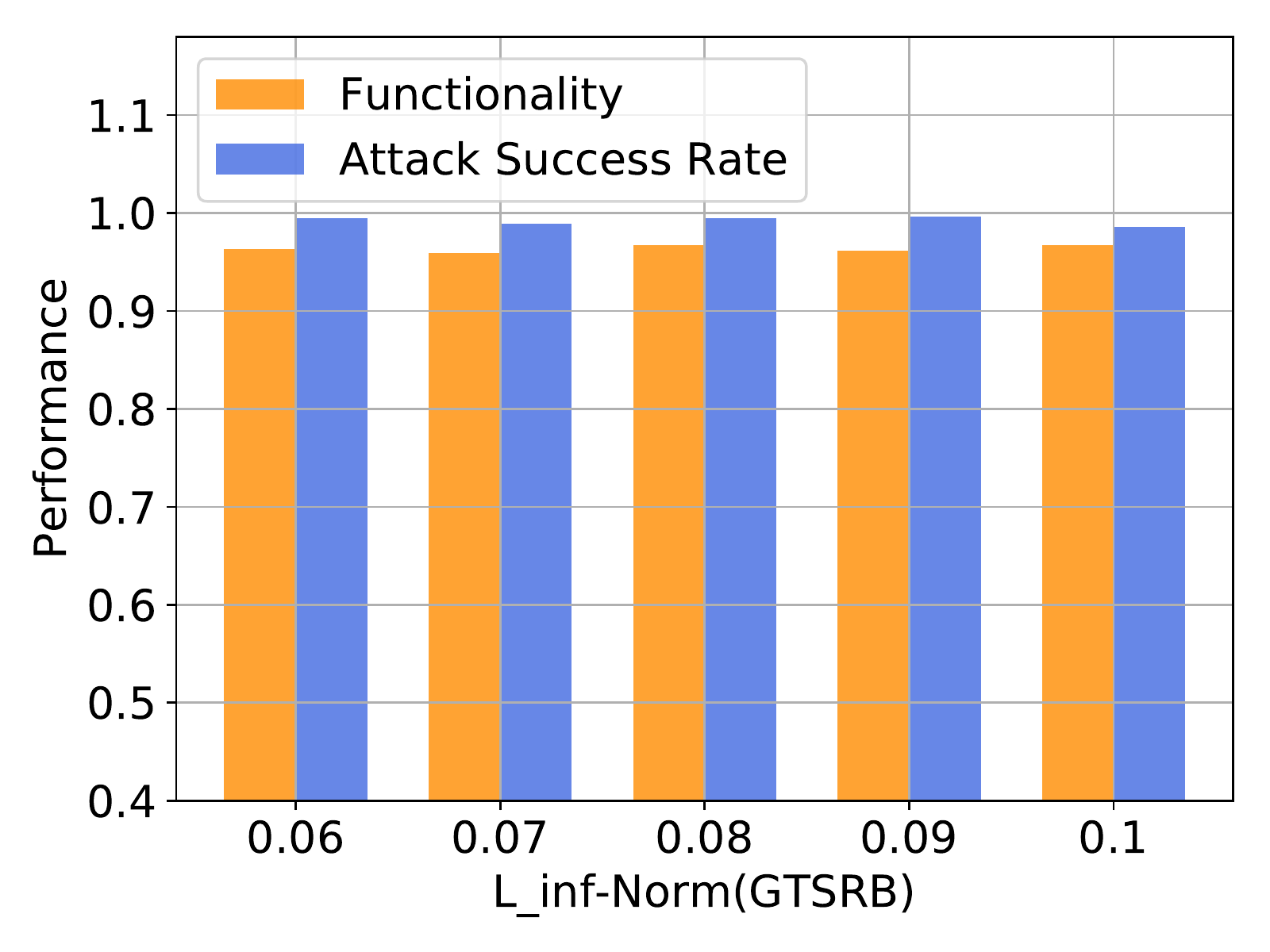}
        \vspace{-0.2cm}
        \label{fig:Li_GT}
    \end{subfigure}
    \vspace{-0.2cm}
    \caption{Functionality and Attack Success Rates of $L_{\infty}$ attack on CIFAR-10/100, GTSRB datasets.}
    \label{fig:li_3ds}
\end{figure*}
For the \textbf{$L_2$-attack}, the results on the CIFAR-10/100 and GTSRB datasets can be seen in Fig.~\ref{fig:l2_3ds}. For the CIFAR-10 dataset, we find that extremely small perturbations ($L_2$-norm $<$ $5$), difficult for humans to perceive, can still produce satisfactory performance in terms of both the \textit{Functionality} and \textit{Attack Success Rate} of the model. The \textit{Attack Success Rate} for all $L_2$-norm tests are greater than $90\%$. For the CIFAR-100 dataset, from Fig.~\ref{fig:l2_3ds} we observe that the \textit{Attack Success Rates} of all the $L_2$ attacks exceed $90\%$. For the GTSRB dataset, we see larger $L_2$-norms are needed on GTSRB to obtain an equivalent \textit{Attack Success Rate} comparable to CIFAR. For instance, only when the $L_2$-norm of the trigger exceeds $9$ does the \textit{Attack Success Rate} exceed $80\%$. With respect to \textit{Functionality}, all configurations retain a validation accuracy comparable to the baseline model, which is 92.48\%, $73.44\%$, and $95.31\%$, respectively.

For the \textbf{$L_0$-attack}, the results can be seen in Fig.~\ref{fig:l0_3ds}. For the CIFAR-100 dataset, with an increase of the $L_0$-norm, the \textit{Attack Success Rate} can be raised to $100\%$. When we retrain the poisoning data on the pre-trained model, we find the model converges faster than the $L_2$-attack to achieve a high \textit{Attack Success Rate}. In only a few epochs, the \textit{Attack Success Rate} exceeds $90\%$, while for $L_2$-norm regularization, it needs more than $10$ epochs to converge. This demonstrates that it is easier for deep neural networks to memorize the triggers generated by $L_0$-norm regularization than $L_2$-norm regularization. It is interesting to see that for the $L_0$-attack, all datasets achieve a higher \textit{Attack Success Rate}. This proves that for those triggers with regular shapes, it is easier for DNNs to identify strong correlated signals, which is also true for human inspectors. With respect to the \textit{Functionality}, all datasets experience a slight drop in the validation accuracy of clean images, but it is acceptable. For example, for CIFAR-100 dataset, the baseline accuracy for CIFAR-100 is $73.44\%$. When compared to the worst configuration ($L_0$-norm $= 1$), the \textit{Functionality} only drops $0.58\%$.

For \textbf{$L_{\infty}$-attack}, we use binary search to find the minimal $L_{\infty}$-norm to achieve over $90\%$ \textit{Attack Success Rate}. We found that when the $L_{\infty}$-norm is less than $0.6$, the optimization for generating the $L_{\infty}$ trigger will get stuck in a sub-optimal oscillation. Thus, we demonstrate $L_{\infty}$ attacks with an $L_{\infty}$-norm beyond $0.6$ as shown in  Fig.~\ref{fig:li_3ds}. We also note that $L_{\infty}$ only penalizes the largest entry of the trigger, resulting in a net change to the input image larger than the $L_2$-attack. 
This large modification (trigger) between the poisoned images and the input images creates more significant effects on the DNN model than the $L_2$ attack, while more invisible for human eyes than the $L_0$ attack.
\begin{table}[t]\footnotesize
    \centering
    \caption{PASS and LPIPS scores compared to the Trojaning attack}
    \vspace{-0.1cm}
    \label{table:2}
    \resizebox{\linewidth}{!}{
    \begin{tabular}{@{}cccccc@{}}
        \toprule
         & Original & \tabincell{c}{Trojan} & \tabincell{c}{$L_2=10$} & \tabincell{c}{$L_0=5$} & \tabincell{c}{$L_{\infty}=0.1$}  \\
        \midrule
        \tabincell{c}{CIFAR-10 \\ \quad } & \includegraphics{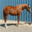} & \includegraphics{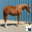} & \includegraphics{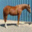} & \includegraphics{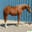} & \includegraphics{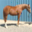} \\
        Avg. PASS & 1 & 0.9610 & 0.9980 & 0.9908 & 0.9900 \\
        Avg. LPIPS & 0.0 & 0.0414 & 0.0156 & 0.0313 & 0.0295 \\
        \midrule
        \tabincell{c}{CIFAR-100 \\ \quad } & \includegraphics{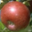} & \includegraphics{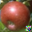} & \includegraphics{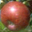} & \includegraphics{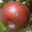} & \includegraphics{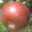} \\
        Avg. PASS & 1 & 0.9543 & 0.9972 & 0.9897 & 0.9685 \\
        Avg. LPIPS & 0.0 & 0.0403 & 0.0134 & 0.0259 & 0.0228 \\
        \midrule
        \tabincell{c}{GTSRB \\ \quad } & \includegraphics{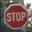} & \includegraphics{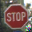} & \includegraphics{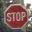} & \includegraphics{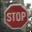} & \includegraphics{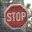}\\
        Avg. PASS & 1 & 0.8920 & 0.9911 & 0.9614 & 0.8904 \\
        Avg. LPIPS & 0.0 & 0.0321 & 0.0062 & 0.0257 & 0.0286 \\
        \bottomrule
    \end{tabular}}
    \vspace{0.5cm}
\end{table}

\noindent \textbf{Comparison with Trojaning Attack.} Table~\ref{tab:cmp_trojan} shows that our attacks achieve fairly commendable performance, with a less obvious mask and a comparable attack success rate when compared to the Trojaning attack.
We further check whether the activation at the neuron position will be affected when applied to various test images. To verify this, we add our $L_p$ trigger into $10000$ test images. Then by observing the average neuron activation of the three backdoored models on the selected position.
We can see from Table~\ref{tab:obser_act} that the average activation at the neuron position is indeed scaled well by our $L_p$-norm trigger, indicating that the attack success is rooted by the scaled neuron activation rather than the trigger itself. This stark difference enables better optimization in teaching the retrained model to recognize these prominent differences as features and use them in predictions. It also makes the attack successful without scaling the neuron activation. 
\begin{table*}[t]
\centering
\caption{Performance in comparison to Trojaning attack}
\label{tab:cmp_trojan}
\resizebox{0.70\textwidth}{!}{
\begin{tabular}{lrrrrrrr}  
\toprule
Attacks  & Target Layer & \tabincell{c}{Initial\\ Activation} & \tabincell{c}{Final\\ Activation} & $L_2$-norm & \tabincell{c}{Poison\\ Rate} & \tabincell{c}{Functionality\\ (\%)} & \tabincell{c}{Attack Success\\ Rate (\%)} \\
\midrule
Trojan & penultimate & 0.3251 & 0.5425 & 23.652 & 0.05 & 91.51 & 100\\
$L_2$ attack    & penultimate & 1.0908 & 3.4893 & 20.000 & 0.05 & 91.02 & 96.44\\
$L_{\infty}$ attack    & penultimate & 0.0132 & 0.2189 & 6.059 & 0.05 & 91.18 & 99.50\\
$L_0$ attack   & penultimate  & 0.3011 & 1.0912 & 24.025 & 0.05 & 91.17 & 99.77\\
\bottomrule
\end{tabular}
}
\end{table*}
\begin{table}[t]
\centering
\caption{Average activation at the selected neuron position}
\label{tab:obser_act}
\begin{tabular}{lccccc} 
\toprule
 Attacks  & \tabincell{c}{Neuron\\ Position} & \tabincell{c}{Trigger\\ Size} & \tabincell{c}{Avg.\\ Clean} & \tabincell{c}{Avg.\\ Attack} \\
\midrule
$L_2$-attack   & 6 & $L_2$-norm=20 & 0.2188 & 0.6288\\
$L_0$-attack  & 6 & $L_0$-norm=25 & 0.3078 & 0.9911\\
$L_{\infty}$-attack & 6 & $L_{\infty}$-norm=0.1 & 0.1975 & 0.7163\\
\bottomrule
\end{tabular}
\end{table}

\begin{figure}[t]
	\begin{subfigure}[t]{0.24\textwidth}
		\centering
		\includegraphics[width=1\linewidth]{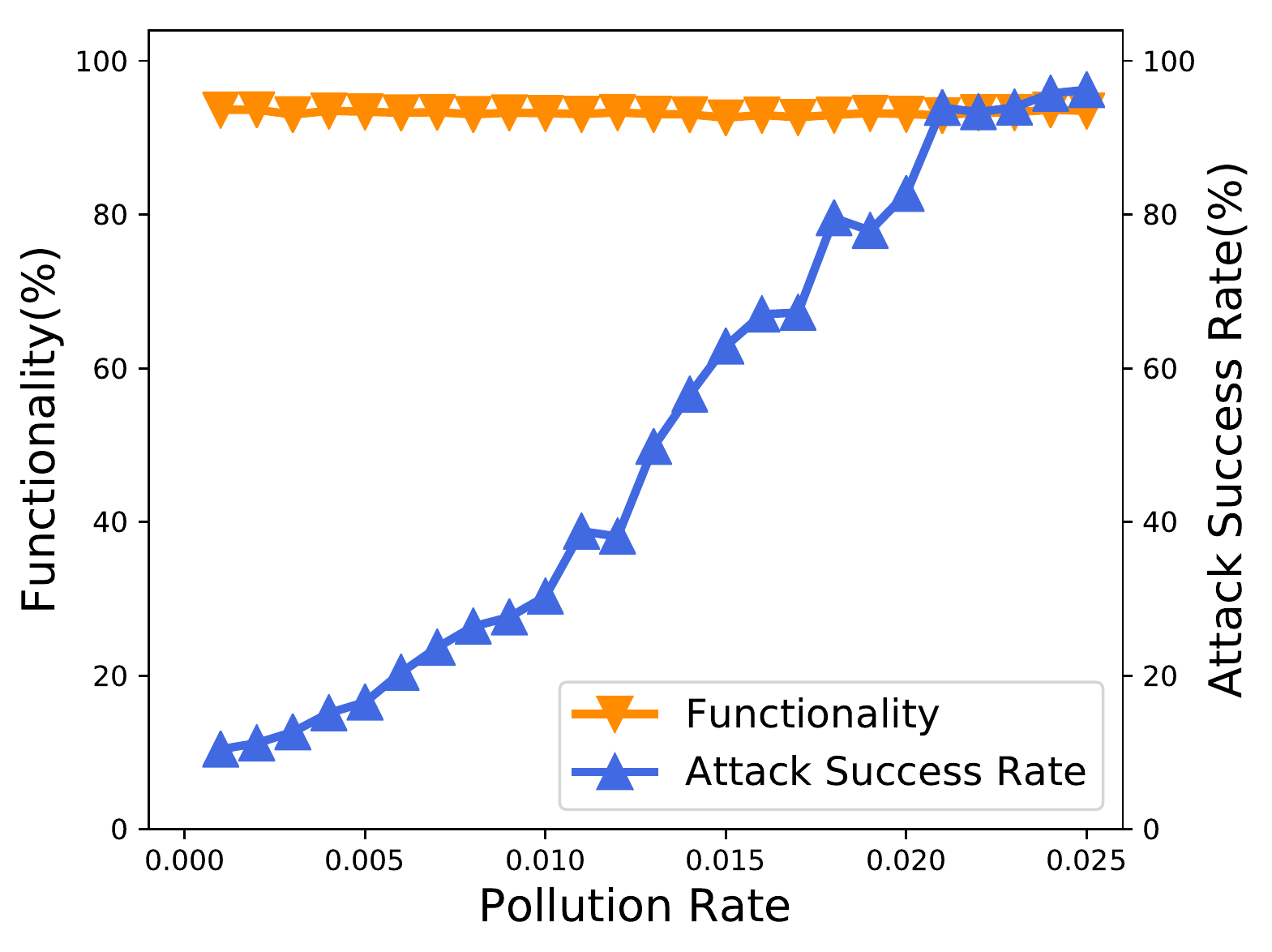}
		\caption{CIFAR-10}
	\end{subfigure}
	\begin{subfigure}[t]{0.24\textwidth}
		\includegraphics[width=1\linewidth]{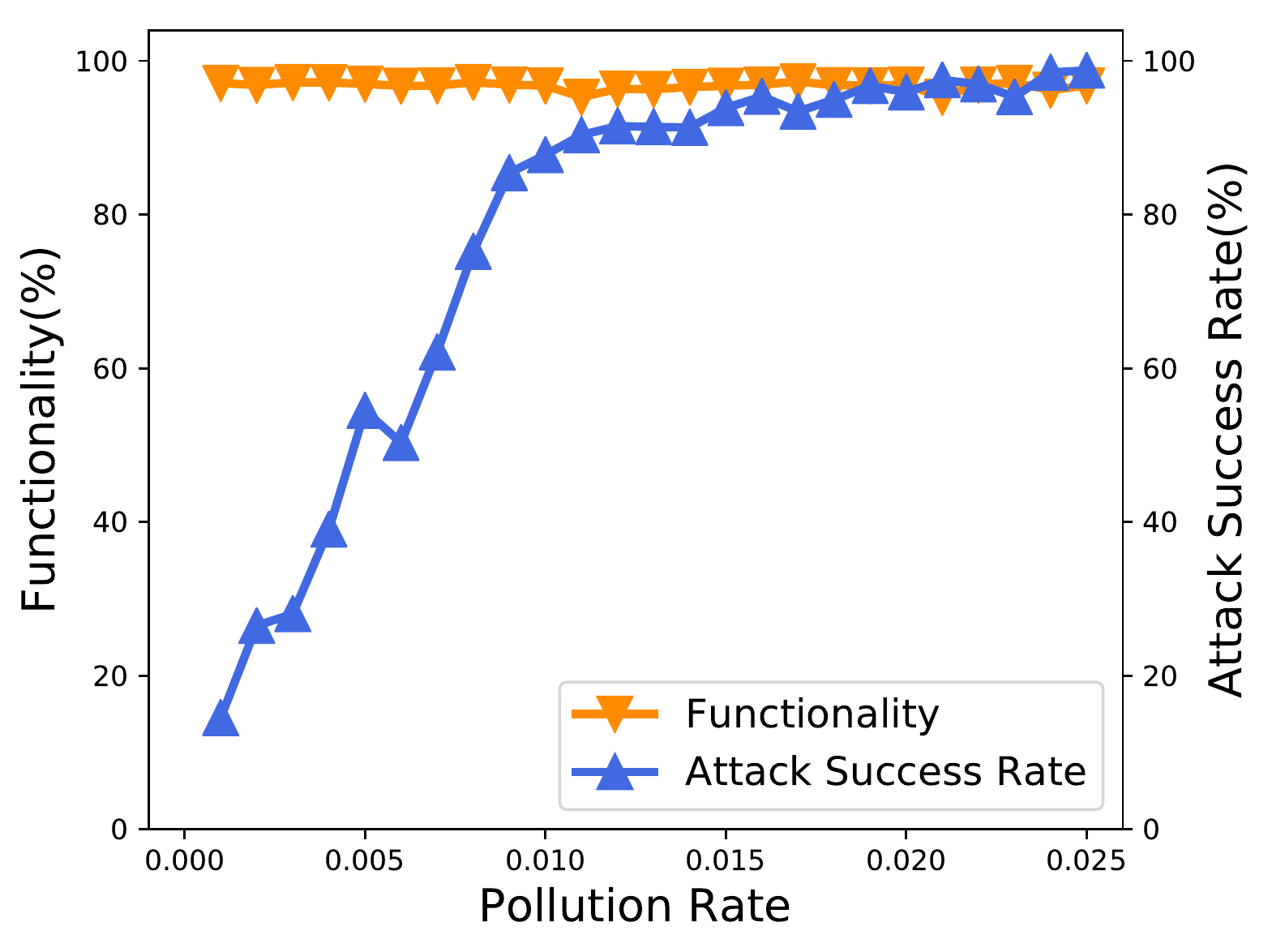}
		\caption{GTSRB}
	\end{subfigure}
	\vspace{-0.2cm}
	\caption{The relationship of \textit{Attack Success Rate} and \textit{Functionality} with the pollution rate $\epsilon$ increase in terms of the $L_{\infty}$ attack on CIFAR-10 (left) and GTSRB (right).}
	\label{fig:prt_li}
\end{figure}

\noindent \textbf{Pollution Rate.} The poisoned samples for the universal attack are drawn from all of the training set, while for the single target attack the poisoned images are only drawn from one source class. In this configuration, we choose the $L_0$ attack ($L_0$-norm $= 5$), the $L_2$ attack ($L_2$-norm = 1) and the $L_{\infty}$ attack ($L_{\infty}$-norm = 0.1) to demonstrate the relationship between the performance of the universal attack and the pollution rate. 
As seen from Fig.~\ref{fig:unv_prt} and Fig.~\ref{fig:prt_li}, with an increasing pollution rate the \textit{Attack Success Rate} increases (the blue line in Fig.~\ref{fig:unv_prt} and Fig.~\ref{fig:prt_li}), while the \textit{Functionality} remains stable. Additionally, for different datasets, the minimal pollution rate required to achieve a high \textit{Attack Success Rate} differs. 

\noindent \textbf{Invisibility Metrics.} Recall that the invisibility metrics are PASS and LPIPS. These metrics quantify how similar two images appear to a human; the range of the PASS metric is $(0,1]$ and for the LPIPS index is $[0,1)$; if two images are identical, the PASS value is $1$ and the LPIPS value is $0$.
We compute and compare the PASS and LPIPS scores between the original image and the poisoning images with triggers generated by Trojaning, $L_2$, $L_0$, and $L_{\infty}$-attacks. The invisibility metrics are shown in Table~\ref{table:2}. Both our triggers achieve a higher average PASS score and a lower average LPIPS score than the Trojaning triggers. Our PASS scores and LPIPS scores are extremely close to $1$ and $0$, respectively. This indicates that humans have more difficulty in discerning differences between the original image and our triggers than the original image and the Trojaning triggers.
The other interesting observation is that the PASS index, a measure of the structural similarity, for the $L_0$-attack is larger than that for the $L_{\infty}$-attack. While the LPIPS index, which not only considers structural similarity but other factors which may affect the perceptual similarity for human,  for the $L_{\infty}$-attack is lower than that for the $L_0$-attack.

\section{Evading Neural Cleanse Detection} \label{sec:nc}
To evaluate the effectiveness of our invisible backdoor attacks to evade the state-of-the-art trojan backdoor detection approaches, we have tested our attacks on \textit{Neural Cleanse}~\cite{wang2019neural}. \textit{Neural Cleanse} operates based on the observation that the minimum perturbation needed to transform inputs of all other source classes to a target class is bounded by the size of the real trigger.
The main part of the \textit{Neural Cleanse} is an optimization framework that generates the minimal perturbation (potential triggers) to misclassify all images drawn from all of other source classes into the target class. The object function shown as Eq.~\eqref{eq:neu_cls}
\begin{equation} \label{eq:neu_cls}
    \min_{m,\Delta} l(y_t, f(A(x, m, \Delta))) + \lambda \cdot \norm{m}_1, 
\end{equation}
where $m$ is the $mask$ which controls the size of the trigger, $\Delta$ is the value of the generated trigger, $y_t$ is the potential target label, and $A(\cdot)$ is the operation adding the trigger into input image. Here the authors of \textit{Neural Cleanse} use the $L_1$-norm to constrain the magnitude of the trigger size. 

The original open sourced \textit{Neural Cleanse} is implemented through Keras on Tensorflow 1.10. We reproduce \textit{Neural Cleanse} with Pytorch 1.4. In \textit{Neural Cleanse}, there are three steps detecting the backdoor attack. First, we need to conduct a backdoor attack to generate a backdoored model to be detected. Note that in their repository, \textit{Neural Cleanse} detects a BadNets backdoored model.  
For comparison, we modify the code to support our four types backdoor attacks (steganography-based, $L_2$, $L_0$, and $L_{\infty}$ regularization-based) with the same network architecture and the same dataset the \textit{Neural Cleanse} repository provides. The parameters and details can be found in Table~\ref{tab:evade_nc_single}, where the first value ``Performance" is the \textit{Functionality} and the second is the \textit{Attack Success Rate}. 
\begin{table}[t]
\centering
\caption{Evaluation the Universal Attacks on Evading Against \textit{Neural Cleanse}}
\label{tab:evade_nc_univ}
\resizebox{0.99\columnwidth}{!}{
\begin{tabular}{lrrrrr}  
\toprule
 Settings & Original NC & $L_2$-attack & $L_0$-attack & $L_{\infty}$-attack \\
\midrule
    Platform     &  Keras 2.2 & Pytorch 1.4 & Pytorch 1.4 & Pytorch 1.4 \\ 
    Network      & 6CNN+2FC & 6CNN+2FC & 6CNN+2FC & 6CNN+2FC     \\
    Dataset      & GTSRB & GTSRB & GTSRB & GTSRB                              \\
    Target Class & [28] & [28] & [28] & [28]                              \\
    Trigger      & 4*4 square & $L_2$=1 & $L_0$=16 & $L_{\infty}$=0.1 \\
    Inject Rate  & 0.1 & 0.1 & 0.1 & 0.1 \\
    Performance  & 0.9678/0.9850 & 0.9445/0.9192 & 0.9684/0.9555 & 0.9668/0.9857 \\
    Detection Result & \{28, 12, 2\} & \{28\} & \{0, 28\} & \{11, 28\} \\
\bottomrule
\end{tabular}
}
\end{table}

For evaluating our attacks on \textit{Neural Cleanse}, firstly we test our three universal backdoor attacks based on $L_p (p=0, 2, \infty)$ regularization. Recall that the universal attack means that there is no restriction to the source input image's class, and any input image with the generated trigger can incur the target label. We report the experimental results in Table~\ref{tab:evade_nc_univ} to evaluate our three regularization based attacks in a universal way against  \textit{Neural Cleanse}. We observe \textit{Neural Cleanse} is able to detect our universal attacks. We believe this is a result of \textit{Neural Cleanse} being able to detect the shortcut of the decision boundary of the backdoored model. For each target class, if there is a minimal perturbation which can transfer all other classes to the target class and the $L_1$-norm of this minimal perturbation is significantly smaller than the minimal perturbations according to other classes, then this class is the target class. An intuitive explanation is that the $L_1$-distance of the shortcut is significantly smaller than other transfer distances, resulting in the anomaly detection algorithm (MAD in \textit{Neural Cleanse}) to find this shortcut.

However, Zhen Xiang et al.~\cite{xiang2019revealing} note that a critical limitation of \textit{Neural Cleanse} is the assumption that the injected backdoor comes from all of other classes. 
Thus, if the attack only uses a source-target pair to inject the backdoor, such as the single target attack we performed in Section~\ref{sec:add_stegan} (Steganography), \textit{Neural Cleanse} will fail to identify the target label from the remaining classes. We further extend our three universal attacks mentioned above to the single target attack. We have experimentally demonstrated this as observed in the Steganography and the extended three regularization based attacks in Table~\ref{tab:evade_nc_single}.
\begin{table}[t]
\centering
\caption{Evaluation the Single Target Attacks on Evading Against \textit{Neural Cleanse}}
\label{tab:evade_nc_single}
\resizebox{0.99\columnwidth}{!}{
\begin{tabular}{lrrrrr}  
\toprule
 Settings & Steganography & $L_2$-attack & $L_0$-attack & $L_{\infty}$-attack \\
\midrule
    Platform     & Pytorch 1.4 & Pytorch 1.4 & Pytorch 1.4 & Pytorch 1.4 \\ 
    Network      & 6CNN+2FC & 6CNN+2FC & 6CNN+2FC & 6CNN+2FC     \\
    Dataset      & GTSRB & GTSRB & GTSRB & GTSRB                              \\
    Source Class & 4 & 4 & 4 & 4                             \\
    Target Class & 7 & 7 & 7 & 7                              \\
    Trigger      & ``Apple``*120 & $L_2$=1 & $L_0$=16 & $L_{\infty}$=0.1 \\
    Inject Rate (Number)  & 1.0(1980) & 1.0(1980) & 0.5(990) & 0.5(990) \\
    Performance  & 0.9492/0.9712 & 0.9550/0.9379 & 0.9536/0.9742 & 0. 9587/0.9545 \\
    Detection Result & \{1,2,0\} & [0, 1, 2\} & \{0, 1\} & \{0, 1, 2\} \\
\bottomrule
\end{tabular}
}
\end{table}
As Table~\ref{tab:evade_nc_single} shows that in our single target attack, the real target is $7$ while the \textit{Neural Cleanse} assumes the target is $\{0, 1, 2\}$, so the \textit{Neural Cleanse} fails to detect the single (source, target) pair attacks.

Following the line of the single (source, target) pair attack, we proposed other methods to evade the detection of  \textit{Neural Cleanse}, which is termed ``Injection All". We make transfer distances from all other classes to each small target class, and make there is no significant difference between transfer distance for each class. As without a clean reference model (a reasonable assumption because the defender only has the provided DNN model in hand~\cite{liu2019abs, guo2019tabor}), the anomaly detection of the \textit{Neural Cleanse} (MAD) will fail. So we inject the backdoor with our invisible backdoor attack for every class using different triggers. We report the experimental results in Table~\ref{tab:evade_nc}.  
The experimental results show that our invisible backdoor attacks can evade the detection of \textit{Neural Cleanse}. We can see that after injecting a backdoor for every class, the maximal anomaly index for each potential target class is below $2$. In Neural Cleanse, if the maximal anomaly index of one class is over $2$, the algorithm will mark this as the target class with a significant confidence. However, as we create trigger shortcuts for each class, their anomaly detection is futile. Additionally, the \textit{Functionality} of the backdoored model is not substantially affected and only decreases slightly. 

\begin{table}[t]
\centering
\caption{Evaluating Against \textit{Neural Cleanse} for Injecting backdoors for All Classes}
\label{tab:evade_nc}
\resizebox{0.99\columnwidth}{!}{
\begin{tabular}{lrrrr}  
\toprule
 Settings & Steganography & $L_2$-attack & $L_0$-attack & $L_{\infty}$-attack \\
\midrule
    Platform     & Pytorch 1.4 & Pytorch 1.4 & Pytorch 1.4 & Pytorch 1.4 \\ 
    Network      & 6CNN+2FC & 6CNN+2FC & 6CNN+2FC & 6CNN+2FC     \\
    Dataset      & GTSRB & GTSRB & GTSRB & GTSRB                              \\
    Target classes & [1-43] & [1-43] & [1-43] & [1-43]        \\
    Trigger Number     & 43 & 43 & 43 & 43 \\
    Inject Rate (Each Class)  & 0.05 & 0.01 & 0.01 & 0.01 \\
    Average Performance & 0.9235/0.9096 & 0.9413/0.9980 & 0.9684/0.9555 & 0.9668/0.9857 \\
    Max Anomaly Index & 1.3742 & 1.7952 & 1.2239  & 1.5652 \\
    Detection Result & \{\} & \{\} & \{\} & \{\} \\
\bottomrule
\end{tabular}
}
\end{table}

According to the optimization framework of \textit{Neural Cleanse} (Eq.~\eqref{eq:neu_cls}), the authors assume that the trigger pattern has a smaller size than the image at the image periphery. In contrast, our steganography-based and $L_2$, $L_{\infty}$ attacks scatter the trigger around the entire image. Therefore, the generated trigger with the optimization framework shown in Eq.~\eqref{eq:neu_cls} greatly differs from our trigger patterns mentioned above. The recovered triggers by \textit{Neural Cleanse} in three different attack settings, the universal attack, the single (source, target) pair attack, and injection all attack can be found in Table~\ref{table:recover}. We acknowledge that Neural Cleanse can still recover our $L_0$ trigger only in the universal setting, which is the primary limitation of our $L_0$ attack.
\begin{table}[t]\footnotesize
    \centering
    \caption{The Recover Trigger of \textit{Neural Cleanse} on Our Invisible Backdoor Attacks by the Single Target Way}
    \vspace{-0.1cm}
    \label{table:recover}
    \resizebox{\linewidth}{!}{
    \begin{tabular}{@{}lcccc@{}}
        \toprule
         & Steganography & \tabincell{c}{$L_2=1$} & \tabincell{c}{$L_0=16$} & \tabincell{c}{$L_{\infty}=0.1$}  \\
        \midrule
        Real Used & \includegraphics{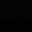} & \includegraphics{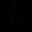} & \includegraphics{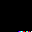} & \includegraphics{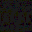} \\
        \midrule
        \tabincell{c}{Recovered\\ (Universal)} & \includegraphics{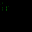} & \includegraphics{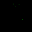} & \includegraphics{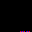} & \includegraphics{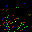} \\
        \midrule
        \tabincell{c}{Recovered\\ (Single Pair)} & \includegraphics{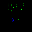} & \includegraphics{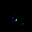} & \includegraphics{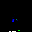} & \includegraphics{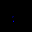} \\
        \midrule
        \tabincell{c}{Recovered\\ (Injection All)} & \includegraphics{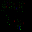} & \includegraphics{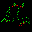} & \includegraphics{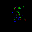} & \includegraphics{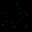} \\
        \bottomrule
    \end{tabular}}
    \vspace{0.5cm}
\end{table}

\section{Discussion on Other Detection Methods} \label{sec:defense}
In this section, we discuss the effectiveness of our backdoor attacks against other backdoor detection approaches. 
The state-of-the-art detection techniques can be categorized into three types: before/during-training, run-time, and post-training.

\noindent{\bf Before/During-Training.} In this scenario, the defender can access the training set (both the poisoning and clean training sets). Before training a DNN model, the defender first checks the training set to identify suspicious training samples, and subsequently removing them. The number of samples in a training set is usually enormous, so prior works~\cite{steinhardt2017certified, statisticalAnalysis} leverage statistical analysis of the poisoned training set to detect whether the training set has been poisoned by a trigger or not. For human inspection before training, we argue that it is challenging to simply perceive the anomaly on our poisoning images~\cite{Liao2018Backdoor} and it is laborious for humans to closely examine such an enormous dataset.

\noindent{\bf Run-Time.} The backdoor detection can also work on the classifier during run-time. In \textit{STRIP}~\cite{gao2019strip}, the authors rely on the strength characteristic of the backdoor attack, which is ``image-agnostic". So the perturbation are added on the inputs to be tested. For trojaned inputs, the predictions of the network is invariant because of the ``image-agnostic" property,  while for clean inputs, when the perturbation is added, the predictions of the network vary dramatically. 
We highlight that \textit{STRIP} is ineffective for defending against our attacks because our irregular triggers are generated with a small perturbation. Therefore, when \textit{STRIP} adds a perturbation to our poisoned inputs, the predictions of the network will also vary greatly due to the fact that our triggers can be broken by adding perturbations. 

\noindent{\bf Post-Training.} The third backdoor detection approach is post-training, where the defender can access the trained models and the partial initial clean training set but with no access to the potential poisoned training set. \textit{Neural Cleanse}~\cite{wang2019neural} and \textit{ABS}~\cite{liu2019abs} are two state-of-the-art model-based white-box defenses. We have discussed \textit{Neural Cleanse} at length in Section~\ref{sec:nc}. On the other hand, the white-box detection method of \textit{ABS}~\cite{liu2019abs}, scans every neuron of the given DNN model, and directly alters the level of stimulation at various neurons; by monitoring the activations of output classes, any neuron which produces a significantly higher output irrespective of the input is an indicator that the resulting model is potentially backdoored. 
Additionally, retraining Trojaned models~\cite{liu2017neural} incurs high computational costs. As prior work has demonstrated~\cite{shan2019gotta, wang2019neural, liu2019abs} that fine-pruning~\cite{liu2018fine} causes the accuracy on unpoisioned data to rapidly plunge when pruning redundant neurons. As for unlearning, there is a condition that the exact trigger must be known~\cite{shan2019gotta, wang2019neural}. In unlearning, they use the reversed trigger to retrain the infected DNN to recognize correct labels even when the trigger is present. As we have shown in Table~\ref{table:recover}, this method cannot recover the real triggers used by our three types of invisible backdoor attacks, and thus these three invisible backdoor attacks are robust to unlearning defenses.

\section{Conclusion}\label{sec:conclusion}
Our work has found that neural networks are sensitive to features imperceptible to humans. We have exploited these features and designed two novel types of backdoor attacks. Because our attack triggers are derived from these covert features, in comparison to the state-of-the-art backdoor attacks, our attack is more covert and overcomes the practicality issues of existing attacks. Because our trigger patterns are invisible to human eyes, non-detection by humans will increase the success probability of backdoor attacks in practice by making the input images inconspicuous. Additionally, we have argued that our triggers can evade the state-of-the-art backdoor detection algorithms, as it is hard to recover the invisible trigger through the optimization algorithm.  

In our future work, we seek to provide a deeper explanation from the internal structure of the neural network to ascertain the reason why the backdoor attack succeeds. In addition, adversarial training~\cite{madry2017towards} and differential privacy~\cite{abadi2016deep,lecuyer2018certified} are proven effective to create a robust machine learning model~\cite{song2019privacy}. We will also explore the feasibility of backdoor attacks against robust models. Understanding the robustness of both attacks and defenses provides an avenue to demystify deep neural networks, rendering deep learning models to be more transparent.

\section*{Acknowledgements}
We would like to thank Jiahao Yu and Damith Ranasinghe for useful discussion and would like to thank Dali Kaafar for proofreading the abstract and introduction of the early version of this work. 
This work was supported, in part, by the National Natural Science Foundation of China, under Grants No. 61972453, No. 61672350, No. U1936214, and No. U1636206.

\ifCLASSOPTIONcaptionsoff
  \newpage
\fi

\bibliographystyle{unsrtnat}\small
\bibliography{tdsc}

\begin{IEEEbiography}[{\includegraphics[width=1.2in,height=1.2in,clip,keepaspectratio]{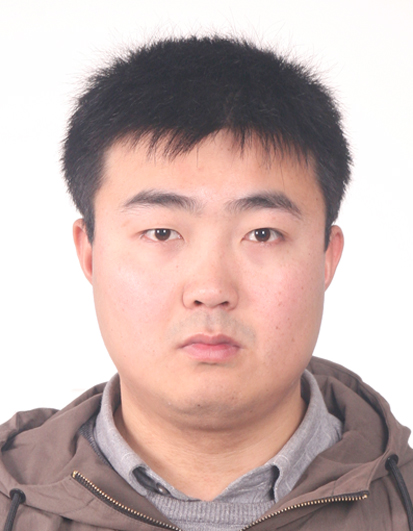}}]
{Shaofeng Li} is pursing his Ph.D. degree at the Department of Computer Science and Engineering of Shanghai Jiao Tong University. He focuses primarily on the areas of machine learning and security, specifically exploring the robustness of machine learning models against various adversarial attacks.   
\end{IEEEbiography}

\begin{IEEEbiography}[{\includegraphics[width=1.2in,height=1.2in,clip,keepaspectratio]{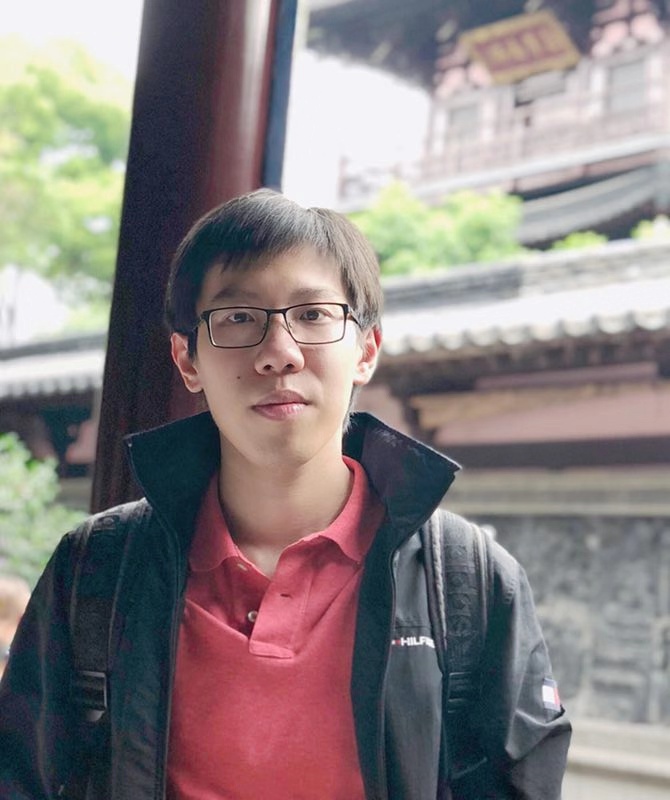}}]
{Minhui Xue} is Lecturer (a.k.a. Assistant Professor) of School of Computer Science at the University of Adelaide. He is also Honorary Lecturer with Macquarie University. His current research interests are machine learning security and privacy, system and software security, and Internet measurement. He published widely in top security and software engineering conferences, including IEEE S\&P, USENIX Security, NDSS, ACM IMC, PETS, IEEE/ACM FSE, IEEE/ACM ASE, and ACM ISSTA. He currently serves on the PC committee of USENIX Security 2021 and PETS 2021. He is the recipient of the ACM SIGSOFT distinguished paper award and IEEE best paper award, and his work has been featured in the mainstream press, including The New York Times and Science Daily. 
\end{IEEEbiography}

\begin{IEEEbiography}[{\includegraphics[width=1.2in,height=1.2in,clip,keepaspectratio]{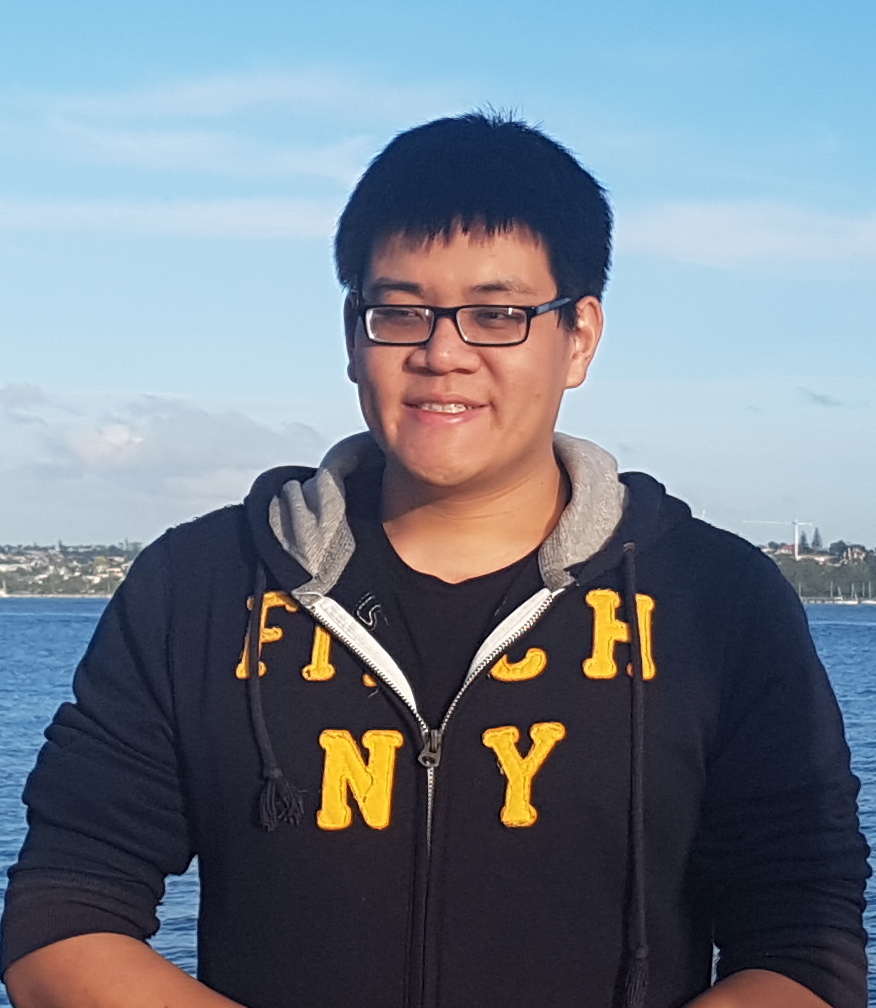}}]
{Benjamin Zi Hao Zhao} is pursuing his Ph.D. at the School of Electrical Engineering and Telecommunications at the University of New South Wales and CSIRO-Data61. His current research interests are authentication systems, and security and privacy with machine learning. His work has received the ACM AsiaCCS best paper award.
\end{IEEEbiography}

\begin{IEEEbiography}[{\includegraphics[width=1in,height=1.25in,clip,keepaspectratio]{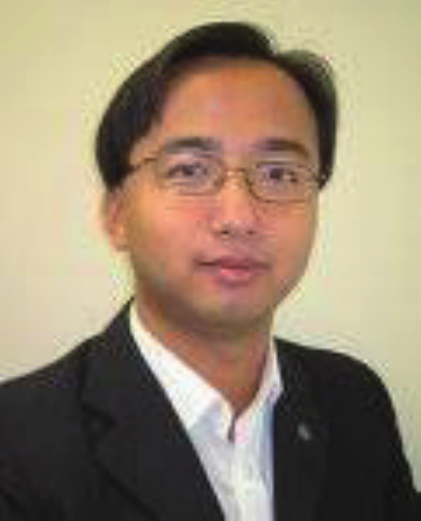}}]{Haojin Zhu} (IEEE M'09-SM'16) received his B.Sc. degree (2002) from Wuhan University (China), his M.Sc. degree (2005) from Shanghai Jiao Tong University (China), both in computer science and the Ph.D. in Electrical and Computer Engineering from the University of Waterloo (Canada), in 2009. Since 2017, he has been a full professor with Computer Science department in Shanghai Jiao Tong University. His current research interests include network security and privacy enhancing technologies. He has published more than 50 journals, including: JSAC, TDSC, TPDS, TMC, TIFS, TWC, TVT and more than 60 international conference papers, including IEEE S\&P, ACM CCS, ACM MOBICOM, NDSS, ACM MOBIHOC, IEEE INFOCOM, IEEE ICDCS. He received a number of awards including: IEEE ComSoc Asia-Pacific Outstanding Young Researcher Award (2014), Top 100 Most Cited Chinese Papers Published in International Journals (2014), Supervisor of Shanghai Excellent Master Thesis Award (2014), Distinguished Member of the IEEE INFOCOM Technical Program Committee (2015), Outstanding Youth Post Expert Award for Shanghai Jiao Tong University (2014), SMC Young Research Award of Shanghai Jiao Tong University (2011). He was a co-recipient of best paper awards of IEEE ICC (2007) and Chinacom (2008), IEEE GLOBECOM Best Paper Nomination (2014), WASA Best Paper Runner-up Award (2017). He received Young Scholar Award of Changjiang Scholar Program by Ministry of Education of P.R. China in 2016. He is a senior member of IEEE and a member of ACM.
 \end{IEEEbiography}
 
\begin{IEEEbiography}[{\includegraphics[width=1in,height=1.25in,clip,keepaspectratio]{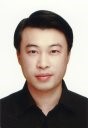}}]{Xinpeng Zhang} received the B.S. degree in computational mathematics from Jilin University, China, in 1995, and the M.E. and Ph.D. degrees in communication and information system from Shanghai University, China, in 2001 and 2004, respectively. Since 2004, he was with the faculty of the School of Communication and Information Engineering, Shanghai University, where he is currently a Professor. He is also with the faculty of the School of Computer Science, Fudan University. He was with The State University of New York at Binghamton as a Visiting Scholar from 2010 to 2011, and also with Konstanz University as an experienced Researcher, sponsored by the Alexander von Humboldt Foundation from 2011 to 2012. His research interests include multimedia security, image processing, and digital forensics. He has published over 200 papers in these areas. He was an Associate Editor of the IEEE TRANSACTIONS ON INFORMATION FORENSICS AND SECURITY from 2014 to 2017.
\end{IEEEbiography}

\end{document}